\documentclass[10pt,twocolumn]{article}
\usepackage[top=0.97in,bottom=0.97in,lmargin=0.75in,rmargin=0.75in]{geometry}
\usepackage[labelfont=bf]{caption}

\usepackage{mathpazo,helvet,inconsolata,bbding,fontawesome}

\usepackage[utf8]{inputenc}
\usepackage[autolanguage,np]{numprint}

\usepackage[english]{babel}
\usepackage{csquotes}
\usepackage{graphicx}
\usepackage{textcomp,gensymb}
\usepackage{amsmath}
\usepackage{amssymb}
\usepackage{xspace}
\usepackage{stackengine}
\usepackage{tabularx}
\usepackage{authblk}

\usepackage[
style=authoryear-comp,
giveninits=true,
uniquename=init,
citestyle=authoryear,
natbib=true,
]{biblatex}
\bibliography{ref}

\bibliography{ref}

\AtEveryBibitem{\clearlist{language}}
\AtEveryBibitem{\clearfield{url}\clearfield{urlyear}}
\AtEveryBibitem{\iffieldundef{doi}{}{\clearfield{url}\clearfield{urlyear}}}
\AtEveryBibitem{\clearfield{month}}
\AtEveryBibitem{\clearlist{publisher}}
\AtEveryBibitem{\clearfield{note}}
\AtEveryBibitem{\clearname{editor}}
\AtEveryBibitem{\clearfield{series}}
\AtEveryBibitem{\clearfield{issn}}
\AtEveryCitekey{\clearfield{issn}}
\renewbibmacro*{url+urldate}{}

\stackMath
\setstackgap{S}{2pt}

\DeclareMathOperator*{\Agg}{\sum_{j=1}^{N}}

\DeclareMathOperator*{\relu}{ReLU}

\newcommand{\loss}{L}

\renewcommand\Vec[1]{\ensuremath{\boldsymbol{#1}}}
\newcommand\Mat[1]{\ensuremath{\boldsymbol{#1}}}

\usepackage[dvipsnames]{xcolor}
\usepackage{hyperref}
\hypersetup{%
	colorlinks=true,%
	linkcolor=BrickRed,%
	citecolor=ForestGreen,
	urlcolor=NavyBlue
}
\usepackage[capitalize,noabbrev,nameinlink]{cleveref}
\crefname{figure}{Figure}{Figures}
\crefname{appendixfigure}{Supplementary Figure}{Supplementary Figures}
\crefname{table}{Table}{Tables}
\crefname{appendixtable}{Supplementary Table}{Supplementary Tables}
\crefname{equation}{Equation}{Equations}
\crefname{appendixequation}{Supplementary Equation}{Supplementary Equations}
\creflabelformat{equation}{#2#1#3}
\creflabelformat{appendixequation}{#2#1#3}

\usepackage{caption}
\title{Graph neural networks uncover structure and functions underlying the activity of simulated neural assemblies}
\author{C\'edric Allier}
\author{Larissa Heinrich}
\author{Magdalena Schneider}
\author{Stephan Saalfeld $^\text{\Envelope}$}
\affil{Janelia Research Campus, Howard Hughes Medical Institute, Ashburn, VA 20147, USA.}
\affil{\href{mailto:saalfelds@janelia.hhmi.org}{\small\Envelope~saalfelds@janelia.hhmi.org \hspace{1cm}\href{https://saalfeldlab.github.io/neural-gnn/}{\faGithub~https://saalfeldlab.github.io/neural-gnn/}}}
\date{\today}

\begin{document}

\twocolumn[ 
\begin{@twocolumnfalse}
\maketitle

\begin{abstract}
\noindent{}Graph neural networks trained to predict observable dynamics can be used to decompose the temporal activity of complex heterogeneous systems into simple, interpretable representations.
Here we apply this framework to simulated neural assemblies with thousands of neurons and demonstrate that it can jointly reveal the connectivity matrix, the neuron types, the signaling functions, and in some cases hidden external stimuli. 
In contrast to existing machine learning approaches such as recurrent neural networks and transformers, which emphasize predictive accuracy but offer limited interpretability, our method provides both reliable forecasts of neural activity and interpretable decomposition of the mechanisms governing large neural assemblies. 
\end{abstract}

\vspace{3em}

\end{@twocolumnfalse} 
]

\section{Introduction}

\begin{figure}
\centering%
\includegraphics[height=\columnwidth-1cm]{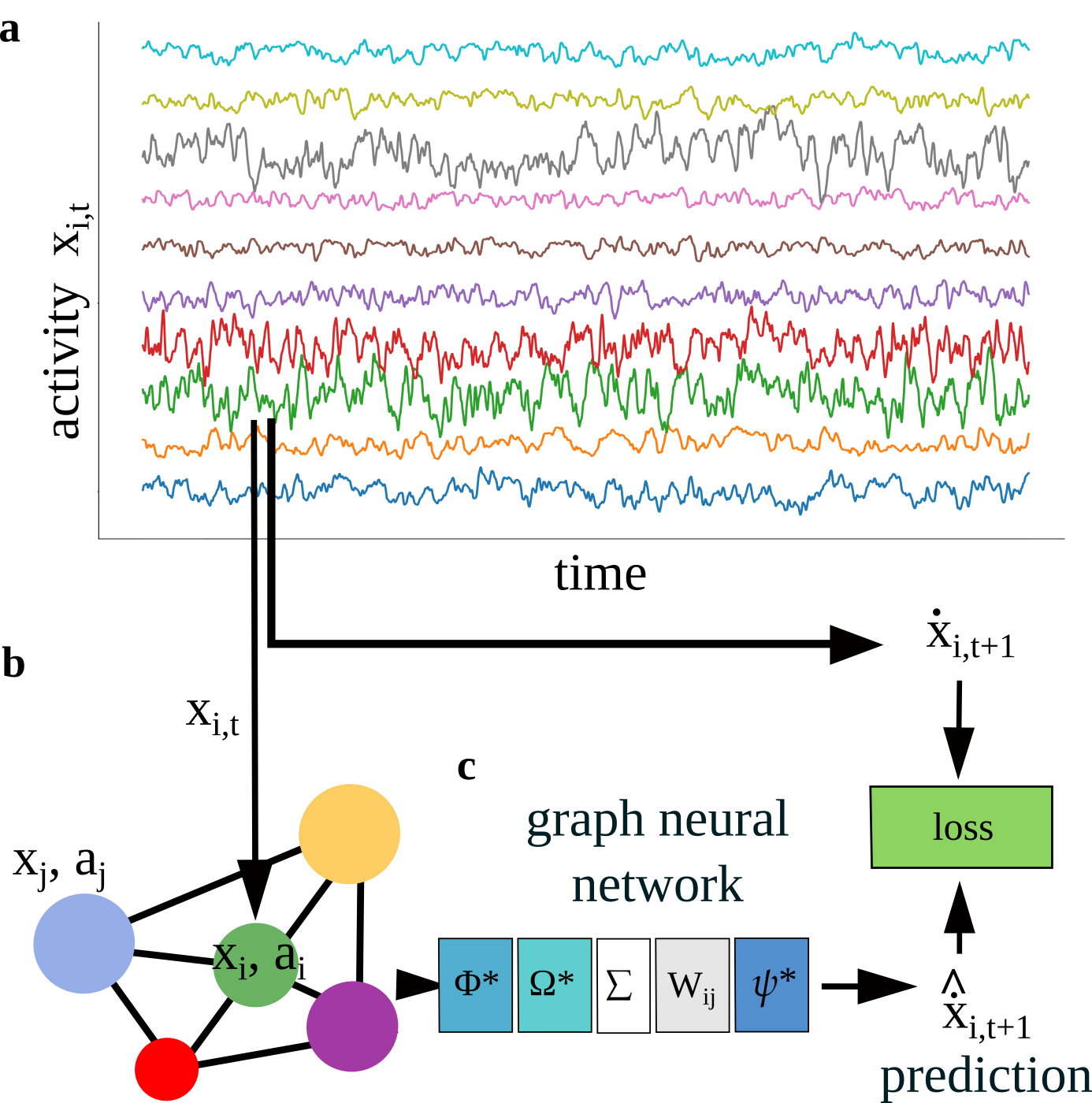}
\caption{The temporal activity of a simulated neural network (\textbf{a}) is converted into densely connected graph (\textbf{b}) processed by a message passing GNN (\textbf{c}). Each neuron (node $i$) receives activity signals $x_j$ from connected neurons (node $j$), processed by a transfer function $\psi^*$ and weighted by the matrix $\Mat{W}$. The sum of these messages is updated with functions $\phi^*$ and $\Omega^*$ to obtain the predicted activity rate $\widehat{\dot{\Vec{x}}}_{i}$.  In addition to the observed activity $x_i$, the GNN has access to learnable latent vectors $\Vec{a}_i$ associated with each node $i$.}
\label{fig:GNN structure}
\end{figure}

We have shown that GNNs can decompose complex dynamical systems into interpretable representations \citep{allier_decomposing_2024}.  We jointly learned pairwise interaction functions and local update rules together with a latent representation of the different objects present in the system, and an implicit representation of external stimuli.  This approach can resolve the complexity arising from heterogeneity in large N-body systems that are affected by complex external inputs.  Here, we leverage this technique to model the simulated activity of large heterogeneous neural assemblies.  We retrieve the connectivity matrix, neuron types, signaling functions, local update rules, and external stimuli from activity data alone, yielding a fully functional mechanistic approximation of the original network with excellent roll-out performance.

Prior studies were successful in retrieving functional connectivity, functional properties of neurons, or predictive models of neural activity at the population level, but a method to infer an interpretable mechanistic model of complex neural assemblies is yet missing. \citet{mi_connectome-constrained_2021} infer the hidden voltage dynamics and functional connectivity from calcium recordings of parts of the \emph{C. elegans} nervous system using a simple biophysical neuron model.  While its predictive performance is modest, it demonstrates that integrating real activity recordings with connectivity can yield experimentally testable predictions beyond simulation.  \citet{pospisil_fly_2024} used the connectome of \emph{Drosophila melanogaster} \citep{dorkenwald_neuronal_2024} to create simulations of full brain activity and inferred effective connectivity from known perturbations of individual neurons with a simple linear model.  Lacking information about connectivity, \citet{wang_foundation_2025} trained a foundation model of mouse visual cortex activity on real two-photon recordings from 135,000 neurons, which predicts responses to both natural videos and out-of-distribution stimuli across animals.
\citet{prasse_predicting_2022} focus on accurate prediction without recovering connectivity while \citet{yu_discovering_2025} infer symbolic dynamical equations without validating connectivity weights. Both work with simulated data and do not learn neuronal heterogeneity.

Our method learns the functional properties of individual neurons and their connections, enabling transfer of the learned dynamics to different network configurations with modified connectivity or different neuron type compositions.

\section{Methods}

\subsection*{Simulation of neural assemblies}

We simulated the activity of neural assemblies with the model described by \citet{stern_reservoir_2023}. We represent the neural system as a graph of $N$ nodes corresponding to neurons and edges representing weighted synaptic connections. The activity signal $x_i$ represents neural dynamics. Each neuron (node~$i$) receives signals from connected neurons (nodes $j$) and updates its own activity $x_i$ according to

\begin{equation}
\dot{x}_i=-\frac{x_i}{\tau_i} + s_i\phi\left(x_i\right) + g_i\Agg \Mat{W}_{ij} \psi_{ij}\left(x_j\right) + \eta_i\left(t\right).
\label{eqn:simulation}
\end{equation}
These systems can generate activity across a wide range of time scales similar to what is observed between cortex regions.  The damping effect (first term) is parameterized by $\tau$, the self-coupling (second term) is parameterized by $s$, and $g$ scales the aggregated messages.  The matrix $\Mat{W}$ contains the synaptic weights multiplying the transfer function $\psi_{ij}(x_j)$.  The weights  were drawn from a Cauchy distribution with \mbox{$\mu=0$} and \mbox{$\sigma^2=\frac{1}{N}$}. A positive and negative sign of the weights indicates excitation and inhibition, respectively.  The last term $\eta_i(t)$ is Gaussian noise with zero mean.  In our experiments, the number of neurons $N$ was \np{1000} or \np{8000}.  We set $g_i$ to 10, and used values between \np{0.25} and \np{8} for $\tau$ and $s$ to test different self-coupling regimes as suggested by \citet{stern_dynamics_2014}.  First, we chose $\tanh(x)$ for both $\phi(x)$ and $\psi_{ij}(x)$.  Later, we made the function $\psi_{ij}$ dependent on the neuron or the interaction between two neurons by changing $\psi_{ij}(x_j)$ to $\tanh(\frac{x_j}{\gamma_i})$ or $\tanh(\frac{x_j}{\gamma_i})-\theta_j x_j$, respectively, with $\gamma$ and $\theta$ parameterizing different neuron types. Finally, we introduced external stimuli into the dynamics through a time-dependent function $\Omega_i(t)$ that scales the aggregated messages.  The model used in our simulations is therefore
\begin{equation}
\begin{split}
\dot{x}_i = & -\frac{x_i}{\tau_i} + s_i\tanh\left(x_i\right) \\
& + g_i\Omega_i\left(t\right)\Agg \Mat{W}_{ij} \left(\tanh\left(\frac{x_j}{\gamma_i}\right) - \theta_j x_j\right) + \eta_i\left(t\right).
\label{eqn:simulation2}
\end{split}
\end{equation}
In \cref{table_sim_parameters}, we list the parameters used for each experiment.

\subsection{Graph neural networks}

\cref{fig:GNN structure} depicts the components of the GNNs trained on simulated data. The GNN learns the update rule
\begin{equation}
\widehat{\dot{\Vec{x}}}_{i} =\phi^*\left(\Vec{a_i}, x_i\right) + \Omega_i^*\left(t\right) \Agg \Mat{W}_{ij}\psi^*\left(\Vec{a_i},\Vec{a_j},x_j\right).
\label{eqn:GNN}
\end{equation}
The optimized neural networks are $\phi^*$, $\psi^*$, modeled as MLPs (ReLU activation, hidden dimension~$=64$, 3 layers, output size~$= 1$), and $\Omega^*$ modeled as a coordinate-based MLP \citep[][, input size~$= 3$,  hidden dimension~$=128$, 5 layers, output size~$= 1$, $\omega=0.3$]{sitzmann_implicit_2020}. Other learnables are the two-dimensional latent vector $\Vec{a}_i$ associated with each neuron, and the connectivity matrix $\Mat{W}$. The optimization loss is
\begin{equation*}
\begin{split}
\loss = {} & \sum_{i = 1}^N \lVert \widehat{\dot{\Vec{x}}}_{i} - \dot{\Vec{x}}_{i}\rVert^2 + \alpha\sum_{i = 1}^N \lVert\phi^*\left(\Vec{a_i}, 0\right)\rVert^2 \\
& + \beta\sum_{i = 1}^N \lVert \relu\left(\frac{\partial\phi^*}{\partial x}\left(\Vec{a_i}, x_i\right)\right)\rVert^2 \\
& + \gamma\sum_{i = 1}^N\sum_{j = 1}^N \lVert \relu\left(-\frac{\partial\psi^*}{\partial x}\left(\Vec{a_i}, \Vec{a_j}, x_j\right)\right)\rVert^2  + \zeta\lVert \Mat{W} \rVert.
\end{split}
\label{eqn:loss_prediction}
\end{equation*}
The first term is the prediction error with $\widehat{\dot{\Vec{x}}}_{i}$ being the GNN prediction, the second term encourages the steady state to be zero, the third term encourages exponential decay to avoid runaway excitations, the fourth term prevents ambiguity about the sign of the connectivity matrix, and the last term encourages sparsity of the connectivity matrix $\Mat{W}$.  In our experiments, we use different combinations of these regularization terms (\cref{table_train_parameters}).  To improve neuron type clustering in the learned latent space, we used a modified version of our heuristic training schedule \citep[][; details in \cref{appendix:cluster-training}]{allier_decomposing_2024}.

We implemented the GNNs using the PyTorch Geometric library \citep{fey_fast_2019} and used AdamUniform gradient descent, with a learning rate of $10^{-4}$. Each GNN was trained over $100$ to $200$ epochs, each epoch covering typically $10^5$ time-points.


\section{Results}

\begin{figure}[t]
\centering%
\includegraphics[width=\columnwidth*8/10]{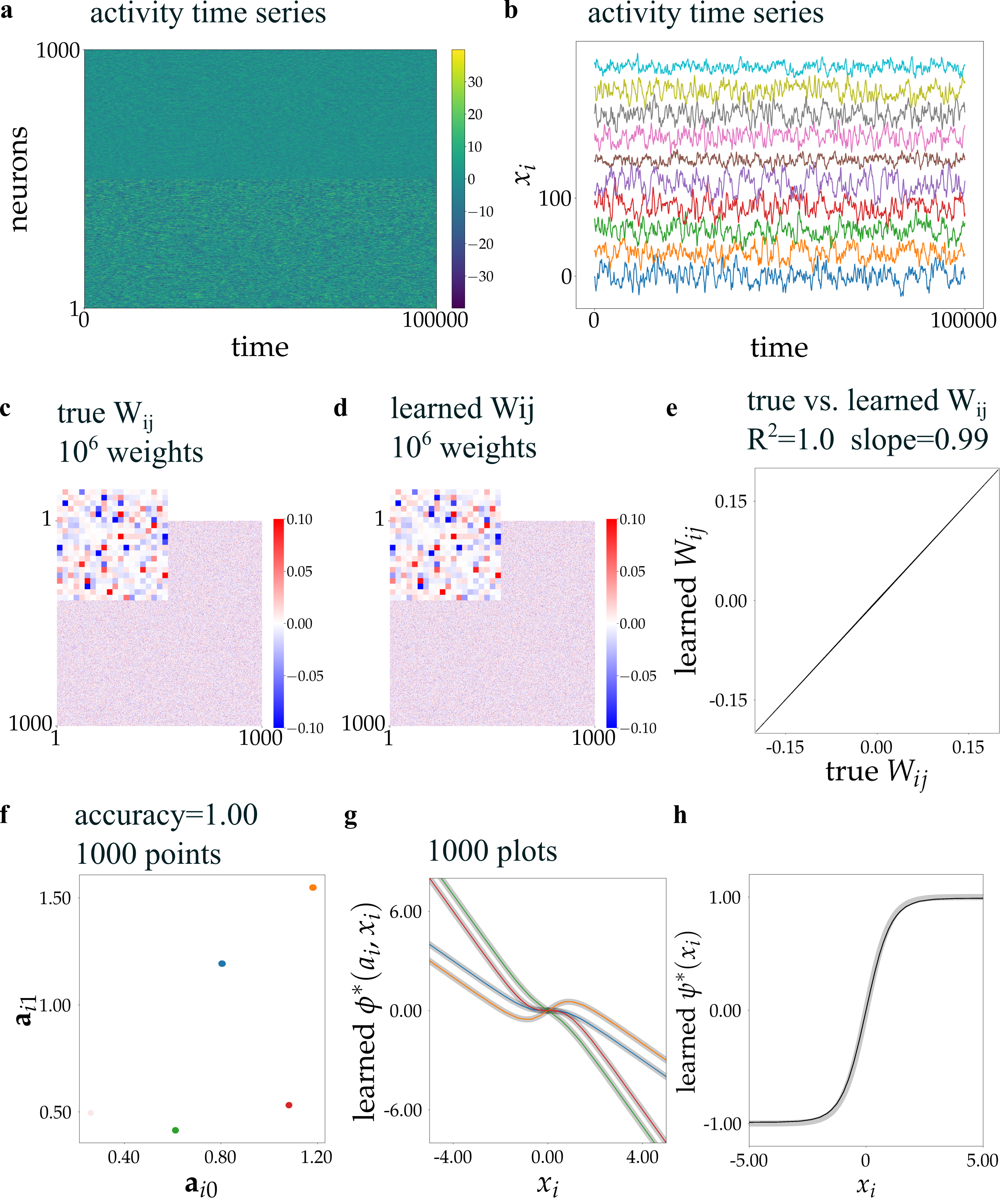}
\caption{\np{1000} densely connected neurons with 4 neuron-dependent update functions. (\textbf{a}) Activity time series used for GNN training. This dataset contains $10^5$ time-points. (\textbf{b}) Sample of 10 time series taken from (\textbf{a}). (\textbf{c}) True connectivity $\Mat{W}_{ij}$. The inset shows $20\times20$ weights. (\textbf{d})~Learned connectivity. (\textbf{e}) Comparison of learned and true connectivity (given $g_i=10$ in \cref{eqn:simulation}). (\textbf{f}) Learned latent vectors $\Vec{a}_i$ of all neurons. (\textbf{g})~Learned update functions $\phi^*(\Vec{a_i}, x_i)$. The plot shows 1000 overlaid curves, one for each vector $\Vec{a_i}$ (\textbf{h}) Learned transfer function $\psi^*(x_i)$, normalized to a maximum value of 1. Colors indicate true neuron types. True functions are overlaid in light gray.}
\label{fig2}
\end{figure}

\begin{figure*}[t]
\centering%
\includegraphics[width=0.715\textwidth]{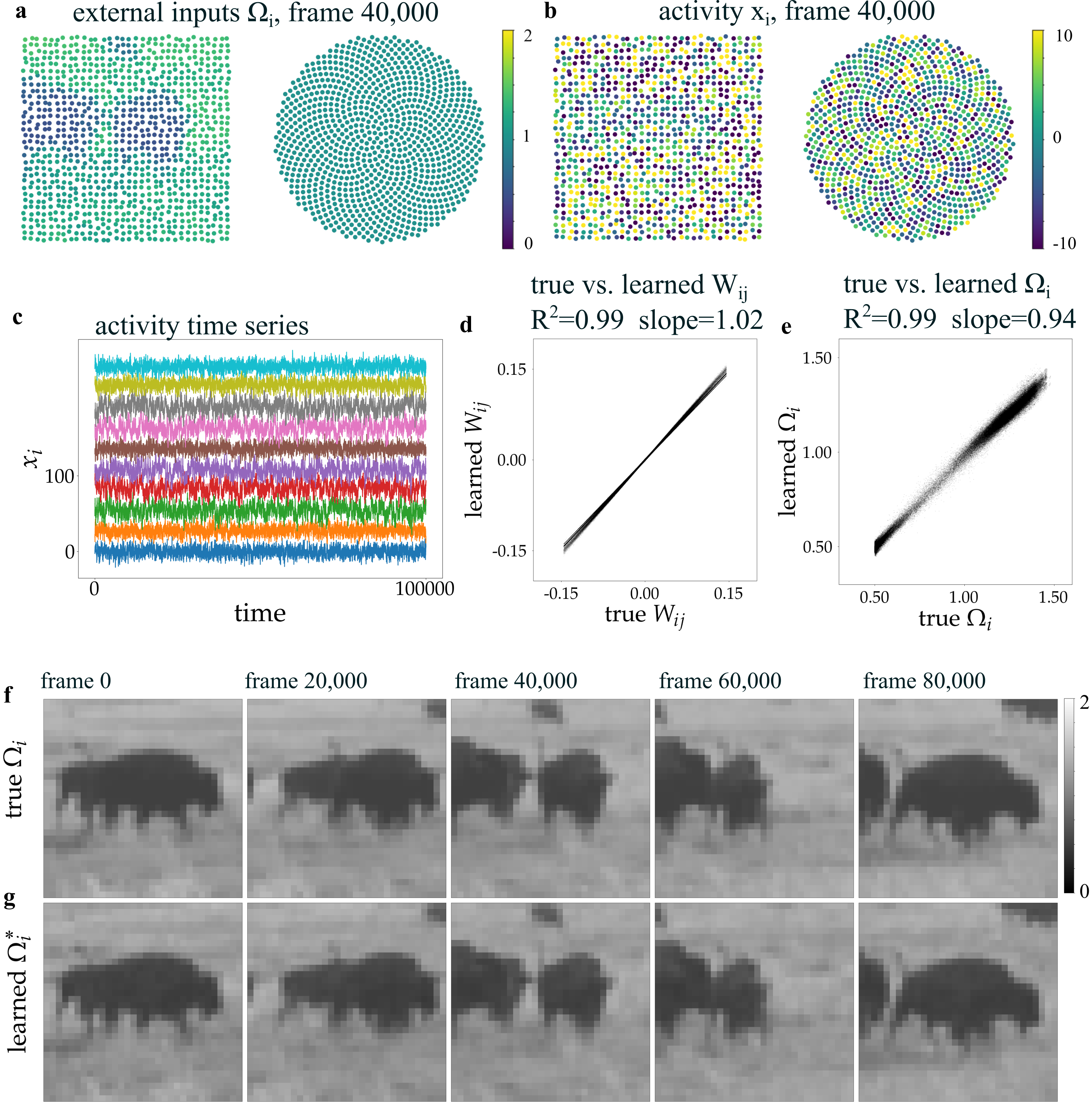}
\caption{\np{2048} densely connected neurons with different neuron-dependent update and transfer functions (4 neuron types), in the presence of external inputs. The training dataset contains $10^5$ time points. (\textbf{a}) External inputs are represented by a time-dependent scalar field $\Omega_i(t)$ that scales the connectivity matrix $\Mat{W}_{ij}$ (\cref{eqn:simulation2}). 1024 neurons (left), spatially ordered, are modulated by this field. The other 1024 neurons (right) are not affected ($\Omega_i=1$). (\textbf{b}) Activity time values. (\textbf{c}) Sample of 10 time series used for training. (\textbf{d}) Comparison of learned and true connectivity  $\Mat{W}_{ij}$ (given $g_i=10$ in \cref{eqn:simulation}). (\textbf{e}) Comparison of learned and true $\Omega_i(t)$ values. (\textbf{f}) True field $\Omega_i(t)$ plotted at different time-points. (\textbf{g})~Learned field $\Omega_i^*(t)$ plotted at different time-points.}
\label{fig3}
\end{figure*}

First, we simulated a noise-free neural network consisting of \np{1000} densely connected neurons of four different types, each parameterized with different values of $\tau_i$ and $s_i$ in \cref{eqn:simulation} for $10^5$ time points. \cref{fig2} shows the simulated training series, and the results of the trained GNN.  It successfully recovered the connectivity matrix ($R^2=1.00$, \mbox{slope~$=0.99$} given $g_i=10$), the common transfer function $\psi$, and the four distinct neuron-specific update functions $\phi^*(\Vec{a})$.   \Cref{supp1} visualizes the joint optimization of shared $\phi*$ and neuron-dependent $\Vec{a}_i$.  Neurons with identical latent parameters (neuron types) eventually form tight clusters of $\Vec{a}$ leading to accurate approximations of the underlying function for each type.

Symbolic regression \citep[PySR package;][]{cranmer_interpretable_2023} over samples generated by $\phi^*$ allowed us to retrieve their exact symbolic expression (\cref{table1}). K-means clustering of the learned latent vectors $\Vec{a}_i$ recovered the neuron types with a classification accuracy of $1.00$. We performed clustering for $K=2,...,10$, and selected the one that yielded the maximum silhouette score (using the Euclidean distance). Increasing the dimension of the latent space did not improve the results, whereas using a one-dimensional latent space was detrimental (data not shown).

To assess generalization capabilities, we performed rollout inference using the trained GNN model with initial activity values not seen during training (\cref{supp12}). The trained GNN predicted the activity of \np{1000} neurons for up to 400 time-points with excellent accuracy ($R^2=0.94$, slope~$=1.00$) and degrades at 800 time-points ($R^2=0.36$, slope~$=0.9$) due to error accumulation.
To assess the importance of learning latent neuron types, we trained a GNN with fixed $\Vec{a}_i$ on the same data (\cref{supp13}) such that the learned update function was no longer neuron-dependent. This GNN recovered the connectivity matrix less well ($R^2=0.92$, slope~$=0.93$) and rollout accuracy was substantially worse (\cref{supp14}).  This suggests that models that ignore the heterogeneity of neural populations are poor approximations of the underlying dynamics of such systems.

Next, we evaluated generalization to fundamentally different network architectures.  We changed the relative proportions of neuron types and the sparsity of the connectivity matrix. The modified GNN model still yielded excellent rollout results (\cref{supp18,supp19}).

Performance scales with the length of the training series.  Systematic sub-sampling of the $10^5$ training series showed that results did not substantially improve beyond $\approx 2\times10^4$ time points for this \np{1000}-neuron system (\cref{supp2}).

Reducing the density of the connectivity matrix leads to an effective reduction of informative training samples.  We therefore tested performance for connectivity matrices with 5\% to 100\% non-zero connections.  Even 5\% connectivity matrices $\Mat{W}$ could be recovered accurately if we enforced sparsity with an L$_1$-penalty on \(\Mat{W}\) (\cref{supp4,supp5}).  If all or some entries known to be zero can be masked, the results are even more robust (data not shown).

To assess robustness, we injected Gaussian noise into the simulated dynamics (\cref{eqn:simulation}) to obtain an SNR of $\sim$10\,dB. At this noise level, training remained stable and we recovered both the connectivity matrix and the signaling functions (\cref{supp6}).

Next, we tested the ability to recover larger connectivity matrices, i.e.\ when the number of neurons was increased to \np{8000}.  \cref{supp3} shows that the 64 million weights of the connectivity matrix in a simulation with Gaussian noise (SNR of $\sim$16\,dB) were well recovered ($R^2=1.00$, slope$=1.00$).  Similarly, increasing the number of neuron types yielded excellent results (\cref{supp7}, 32 different update functions, classification accuracy of $0.99$).

We then introduced neuron-specific transfer functions of the form $\psi\!\big(x_j/\gamma_i\big)$ (\cref{supp8}). Jointly optimizing the shared MLP $\psi^*$ and the latent vectors $\Vec{a}_i$ accurately identified these functions; symbolic regression recovered their analytical expressions (\cref{table2}). The learned connectivity matched the ground truth ($R^2=0.99$, slope~$=0.99$).

Next, we examined neuron–neuron–specific transfer functions $\psi\!\big(x_j/\gamma_i-\theta_i x_j\big)$ (\cref{supp9}).  Here, the multivariate MLP $\psi^*$ takes pairwise latents $(\Vec{a_i},\Vec{a_j})$ as inputs (\cref{table_train_parameters}).  The four neuron types were well recovered, as were the four corresponding update functions and the 16 pairwise transfer functions.  Symbolic regression yielded close approximations (\cref{table3}). The learned connectivity again matched the ground truth
($R^2=0.99$, slope$=1.03$).

As neural networks in nature do not work in isolation but process external inputs, we tested whether we could recover both network structure and dynamics, as well as unknown external inputs in a final experiment.  We introduced a time evolving function $\Omega_i(t)$ in the simulation (\cref{eqn:simulation2}), being spatially defined on a grid of \np{1024} neurons (\cref{fig3}).  During training, $\Omega_i(t)$ is approximated by a coordinate-based MLP $\Omega^*(x, y, t)$.  In addition to the grid of \np{1024} neurons impacted by external stimuli, the simulation includes another set of \np{1024} neurons for which $\Omega_i=1$. The external inputs were well recovered ($R^2=0.99$, slope$=1.02$), together with the connectivity matrix ($R^2=0.99$, slope$=0.94$), neuron types, and signaling functions (\cref{supp10}).

\section{Discussion}

To enable future applications to experimental data that captures neural activity of large complex networks \citep{lueckmann_zapbench_2025}, we designed simulations of complex neural assemblies.  We covered systems with diverse functional connectivity, different neuron types affecting signal transfer between neurons and internal state update, and external inputs.  With these simulations, we showed that message-passing GNNs can be used to recover the structure, functions, and the external inputs underlying the observed activity.

Our GNN-based approach is immediately interpretable, because it models the inputs, outputs, and behavior of individual neurons, and their functional connectivity.  In our experiments, the learned latent embedding of individual neurons is two-dimensional, offering an excellent visualization that can be used to analyze the structure of the underlying parameterization (e.g.\ discrete neuron types vs.\ continuous parameterization).  The signaling functions are approximated by simple MLPs that can be used as sample generators for symbolic regression to recover their analytical forms.

We showed that the joint optimization of simple shared MLPs and low-dimensional latent embeddings for individual neurons is a powerful tool to account for the variability present in neural assemblies.  It allows to distinguish different types of neurons, to reveal their signaling functions, and unobserved external inputs.

GNNs are an excellent tool to model known properties of the structure and function of complex dynamical systems such as neural networks.  If partial knowledge about the connectivity of neurons is available, the connectivity matrix can be masked, neurons known to be of the same type can be forced to share learnable embeddings, expectations about the structure of the signaling and update functions can be used to constrain what the model can learn.

While our simulations capture core features of neural dynamics, biological applications will require additional features. Therefore, in future work, we will add time-dependent connectivity, signaling functions, and neural embeddings, time delays, diffusive exchange of chemical signals, and the fact that the neural dynamics are not directly observed but the result of a poorly characterized indirect biophysical process.

\section{Conclusion}

Our approach demonstrates the potential of GNNs to model complex neural activity, recovering key components such as connectivity matrices, signaling functions, and external inputs from observed dynamics alone.  The method effectively decomposes the observed complexity into simple, interpretable components. Future work will focus on incorporating additional key features of neural activity, such as dynamically changing connectivity, time delays, and neuromodulation.



\section*{Acknowledgements}

We thank Tosif Ahamed and Tirthabir Biswas for helpful discussions and comments that improved this work and manuscript, and Michael Innerberger for making the code more accessible. 

\printbibliography

\appendix

\renewcommand{\figurename}{Supplementary Figure}
\setcounter{figure}{0}
\crefalias{figure}{appendixfigure}%
\newcommand{\equationname}{Supplementary Equation}
\setcounter{equation}{0}
\crefalias{equation}{appendixequation}%
\renewcommand{\tablename}{Supplementary Table}
\setcounter{table}{0}
\crefalias{table}{appendixtable}%

\clearpage

\section{Supplementary notes}

\subsection{Neuron type clustering used during training}
\label{appendix:cluster-training}

In the main training loop, we jointly train the parameters of the MLPs $\phi^*$ and $\psi^*$, the coordinate-based network $\Omega^*$, and the latent vectors $\Vec{a}_i$ for each neuron.  This does not guarantee that similar learned functions $\phi^*$ and $\psi^*$ are produced by similar latent vectors $\Vec{a}_i$, and vice versa.  To encourage such a  well-behaved mapping, we repeatedly cluster and re-initialize the vectors $\Vec{a}$ for all neurons and retrain the corresponding function $\phi^*$.  We perform this clustering every 4 epochs of training:

\begin{description}
\item[Step 1:]{Sample function profiles
\[
\Vec{F}_i = \phi^*(\Vec{a}_i, x_j), \quad x_j \in [-5,5], \quad 1000~\text{samples}
\]
}
\item[Step 2:]{Project profiles to 2D with UMAP
\[
\Vec{z}_i = \text{UMAP}(\Vec{F}_i) \in \mathbb{R}^2
\]
}
\item[Step 3:]{Hierarchical clustering $c$ of $\Vec{z}_i$ (complete linkage with Euclidean distance threshold 0.1), resulting in $m$ clusters

\[
C_k = \{i: c(i) = k\}
\]
}
\item[Step 4:]{Define target functions
\[
f^*_k(x) = \underset{i \in C_k}{\text{median}}(\phi^*(\Vec{a}_i, x))
\]
}
\item[Step 5:]{Replace latent vectors with cluster medians
\[
\Vec{a}'_k = \underset{i \in C_k}{\text{median}}(\Vec{a}_i), \quad \Vec{a}_i \leftarrow \Vec{a}'_k \quad \forall i \in C_k
\]
}
\item[Step 6:]{Retrain $\phi^*$ with loss
\[
\sum_{k=1}^m \left\| \phi^*(\Vec{a}'_k, x) - f_k^*(x) \right\|^2
\]
for 20 epochs of 1000 samples $x \in [-5,5]$.
}
\end{description}

\clearpage
\onecolumn
\noindent
The model used in our simulations is
\begin{equation}
\begin{split}
\dot{x}_i = {} & -\frac{x_i}{\tau_i} + s_i\tanh(x_i) \\
& + g_i\Omega_i(t)\Agg \Mat{W}_{ij} (\tanh(\frac{x_j}{\gamma_i}) - \theta_j x_j) + \eta_i(t).
\label{eqn:simulation3}
\end{split}
\end{equation}
\noindent
Following table summarizes the simulation parameters.
\begin{table*}[h]
\centering
\footnotesize
\begin{tabular}{|l|p{1.6cm}|c|c|c|c|c|c|c|c|c|c|c|c|}
\hline
\textbf{Name} & \textbf{Figure} & $\mathbf{N_{\text{frames}}}$ & $\mathbf{N_{\text{neurons}}}$ & $\mathbf{N_{\text{types}}}$ & \textbf{Conn.} & $\boldsymbol{\sigma^2}$ & $\boldsymbol{\Omega}$ & $\mathbf{g_i}$ & $\mathbf{s_i}$ & $\boldsymbol{\tau_i}$ & $\boldsymbol{\gamma_j}$ & $\boldsymbol{\theta_j}$ \\
\hline
Baseline & 2, Supp. 1-7 & \np{100000} & \np{1000} & 4 & 100\% & 0 & no & 10 & 1,2 & 0.5,1 & 1 & 0 \\
\hline
External inputs & 3, Supp. 15 & \np{100000} & \np{2048} & 4 & 100\% & 1 & yes & 10 & 1,2 & 0.5,1 & 1,2,4,8 & 0 \\
\hline
Sparse & Supp. 8-9 & \np{100000} & \np{1000} & 4 & 5\% & 0 & no & 10 & 1,2 & 0.5,1 & 1 & 0 \\
\hline
High noise & Supp. 10 & \np{100000} & \np{1000} & 4 & 100\% & 7.2 & no & 10 & 1,2 & 0.5,1 & 1 & 0 \\
\hline
Large scale & Supp. 11 & \np{100000} & \np{8000} & 4 & 100\% & 1 & no & 10 & 1,2 & 0.5,1 & 1 & 0 \\
\hline
Many types & Supp. 12 & \np{100000} & \np{1000} & 32 & 100\% & 0 & no & 10 & 1-8 & 0.25-1 & 1 & 0 \\
\hline
Transmitters & Supp. 13 & \np{100000} & \np{1000} & 4 & 100\% & 0 & no & 10 & 1,2 & 0.5,1 & 1,2,4,8 & 0 \\
\hline
Transmitters \& receptors & Supp. 14 & \np{100000} & \np{1000} & 4 & 100\% & 0 & no & 10 & 1,2 & 0.5,1 & 1,2,4,8 & 0-0.040 \\
\hline
\end{tabular}
\caption{Simulation parameters. Connectivity indicates percentage of non-zero $W_{ij}$. Noise $\sigma^2$ is variance of $\eta_i(t)$ in \cref{eqn:simulation3}. $\Omega$ indicates presence of external inputs $\Omega_i(t)$. Parameters: $g_i$ (coupling strength), $s_i$ (self-coupling), $\tau_i$ (time constant), $\gamma_j$ (scale in $\tanh(x_j/\gamma_i)$), $\theta_j$ (linear term in $\tanh(x_j/\gamma_i)-\theta_j x_j$). }
\label{table_sim_parameters}
\end{table*}

\noindent
The GNN learns the update rule
\begin{equation}
\widehat{\dot{\Vec{x}}}_{i} =\phi^*(\Vec{a_i}, x_i) + \Omega_i^*(t) \Agg \Mat{W}_{ij}\psi^*(\Vec{a_i},\Vec{a_j},x_j).
\label{eqn:GNN2}
\end{equation}
The optimized neural networks are $\phi^*$, $\psi^*$, modeled as MLPs (ReLU activation, hidden dimension~$=64$, 3 layers, output size~$= 1$), and $\Omega^*$ modeled as a coordinate-based MLP \citep[][, input size~$= 3$,  hidden dimension~$=128$, 5 layers, output size~$= 1$, $\omega=0.3$]{sitzmann_implicit_2020}. Other learnables are the two-dimensional latent vector $\Vec{a}_i$ associated with each neuron, and the connectivity matrix $\Mat{W}$. The optimization loss is
\begin{equation}
\begin{split}
\loss = {} & \sum_{i = 1}^N \lVert \widehat{\dot{\Vec{x}}}_{i} - \dot{\Vec{x}}_{i}\rVert^2 + \alpha\sum_{i = 1}^N \lVert\phi^*(\Vec{a_i}, 0)\rVert^2 \\
& + \beta\sum_{i = 1}^N \lVert \relu(\frac{\partial\phi^*}{\partial x}(\Vec{a_i}, x_i))\rVert^2 \\
& + \gamma\sum_{i = 1}^N\sum_{j = 1}^N \lVert \relu(-\frac{\partial\psi^*}{\partial x}(\Vec{a_i}, \Vec{a_j}, x_j))\rVert^2  + \zeta\lVert \Mat{W} \rVert.
\end{split}
\label{eqn:loss_prediction2}
\end{equation}
\noindent
Following table summarizes the training parameters.

\begin{table*}[h]
\centering
\footnotesize
\begin{tabular}{|l|p{2.2cm}|c|c|c|c|c|}
\hline
\textbf{Name} & \textbf{Figure} & $\boldsymbol{\alpha}$ & $\boldsymbol{\beta}$ & $\boldsymbol{\gamma}$ & $\boldsymbol{\zeta}$ & $\boldsymbol{\psi^*}$ \textbf{input} \\
\hline
Baseline & 2, Supp. 1-7 & 1 & 0 & 10 & 0 & $x_j$ \\
\hline
External inputs & 3, Supp. 15 & 1 & 5 & 10 & $10^{-5}$ & $a_j, x_j$ \\
\hline
Sparse & Supp. 8-9 & 1 & 0 & 10 & $10^{-5}$ & $x_j$ \\
\hline
High noise & Supp. 10 & 1 & 0 & 10 & 0 & $x_j$ \\
\hline
Large scale & Supp. 11 & 1 & 0 & 10 & $5 \cdot 10^{-5}$ & $x_j$ \\
\hline
Many types & Supp. 12 & 1 & 0 & 10 & 0 & $x_j$ \\
\hline
Transmitters & Supp. 13 & 1 & 0 & 100 & 0 & $a_j, x_j$ \\
\hline
Transmitters \& receptors & Supp. 14 & 1 & 0 & 500 & 0 & $a_i, a_j, x_j$ \\
\hline
\end{tabular}
\caption{Training parameters for loss function (\cref{eqn:loss_prediction2}).}
\label{table_train_parameters}
\end{table*}

\clearpage
\twocolumn

\begin{table}[h]
\centering
\footnotesize
\begin{tabular}{|c|c|c|}
\hline
\textbf{function} & \textbf{true} & \textbf{learned} \\
\hline
$\phi_1$ & $-x + \tanh (x)$ & $-0.998x +\tanh (x) - 0.0016$ \\
\hline
$\phi_2$ & $-x + 2\tanh (x)$ & $-0.998x +1.996\tanh (x)$ \\
\hline
$\phi_3$ & $-2x + \tanh (x)$ & $-1.994x +\tanh (x)$ \\
\hline
$\phi_3$ & $-2x + 2\tanh (x)$ & $-1.996x +1.997\tanh (x)$ \\
\hline
$\psi$ & $\tanh (x)$ & $\tanh (x)$ \\
\hline
\end{tabular}
\caption{Comparison of true and learned functions plotted in \cref{fig2}. Symbolic regression (PySR package \citep{cranmer_interpretable_2023}) is applied to the learned functions to retrieve their expressions.}
\label{table1}
\end{table}

\begin{table}[h]
\centering
\footnotesize
\begin{tabular}{|c|c|c|}
\hline
\textbf{function} & \textbf{true} & \textbf{learned} \\
\hline
$\phi_1$ & $-x + \tanh (x)$ & $-0.999x +\tanh (x)$ \\
\hline
$\phi_2$ & $-x + 2\tanh (x)$ & $-0.999x +1.992\tanh (x)$ \\
\hline
$\phi_3$ & $-2x + \tanh (x)$ & $-1.994x +\tanh (x)$ \\
\hline
$\phi_3$ & $-2x + 2\tanh (x)$ & $-1.984x +1.991\tanh (x)$ \\
\hline
$\psi_1$ & $\tanh (x)$ & $\tanh (x)$ \\
\hline
$\psi_2$ & $\tanh (0.5x)$ & $\tanh (0.489)x$ \\
\hline
$\psi_3$ & $\tanh (0.25x)$ & $\tanh (0.247x)$ \\
\hline
$\psi_4$ & $\tanh (0.125x)$ & $\tanh (0.128x)$ \\
\hline
\end{tabular}
\caption{Comparison of true and learned functions plotted in \cref{supp8}.}
\label{table2}
\end{table}

\begin{table}[h]
\centering
\footnotesize
\begin{tabular}{|c|c|c|}
\hline
\textbf{function} & \textbf{true} & \textbf{learned} \\
\hline
$\phi_1$ & $-x + \tanh (x)$ & $-0.999x + \tanh (x)$ \\
\hline
$\phi_2$ & $-x + 2\tanh (x)$ & $-0.993x +1.999\tanh (x)$\\
\hline
$\phi_3$ & $-2x + \tanh (x)$ & $-1.992x +\tanh (x)$ \\
\hline
$\phi_3$ & $-2x + 2\tanh (x)$ & $-1.974x +1.989\tanh (x)$ \\
\hline
$\psi_{11}$ & $\tanh (x)$ & $\tanh (x)$ \\
\hline
$\psi_{12}$ & $\tanh (x) - 0.013 x$ & $\tanh (x) - 0.017x$ \\
\hline
$\psi_{13}$ & $\tanh (x) - 0.027 x$ & $\tanh (x) -0.028x$ \\
\hline
$\psi_{14}$ & $\tanh (x) - 0.040 x$ & $\tanh (x) -0.053x$ \\
\hline
$\psi_{21}$ & $\tanh (0.5x)$ & $\tanh (0.486)x$ \\
\hline
$\psi_{22}$ & $\tanh (0.5x) - 0.013 x$ & $\tanh(0.414x)$ \\
\hline
$\psi_{23}$ & $\tanh (0.5x) - 0.027 x$ & $0.814 \tanh (0.603x)$ \\
\hline
$\psi_{24}$ & $\tanh (0.5x) - 0.040 x$ & $0.67 \tanh (x)$ \\
\hline
$\psi_{31}$ & $\tanh (0.25x)$ & $\tanh (0.222x)$ \\
\hline
$\psi_{32}$ & $\tanh (0.25x) - 0.013 x$ & $\tanh (0.204x)$ \\
\hline
$\psi_{33}$ & $\tanh (0.25x) - 0.027 x$ & $\tanh (0.163x)$ \\
\hline
$\psi_{34}$ & $\tanh (0.25x) - 0.040 x$ & $\tanh (0.172x)$ \\
\hline
$\psi_{41}$ & $\tanh (0.125x)$ & $\tanh (0.118x)$ \\
\hline
$\psi_{42}$ & $\tanh (0.125x) - 0.013 x$ & $\tanh (0.110x)$ \\
\hline
$\psi_{43}$ & $\tanh (0.125x) - 0.027 x$ & $\tanh (0.097x)$ \\
\hline
$\psi_{44}$ & $\tanh (0.125x) - 0.040 x$ & $\tanh (0.081x)$ \\
\hline
\end{tabular}
\caption{Comparison of true and learned functions plotted in \cref{supp9}.}
\label{table3}
\end{table}

\begin{figure*}[h]
\centering%
\includegraphics[width=0.9\textwidth]{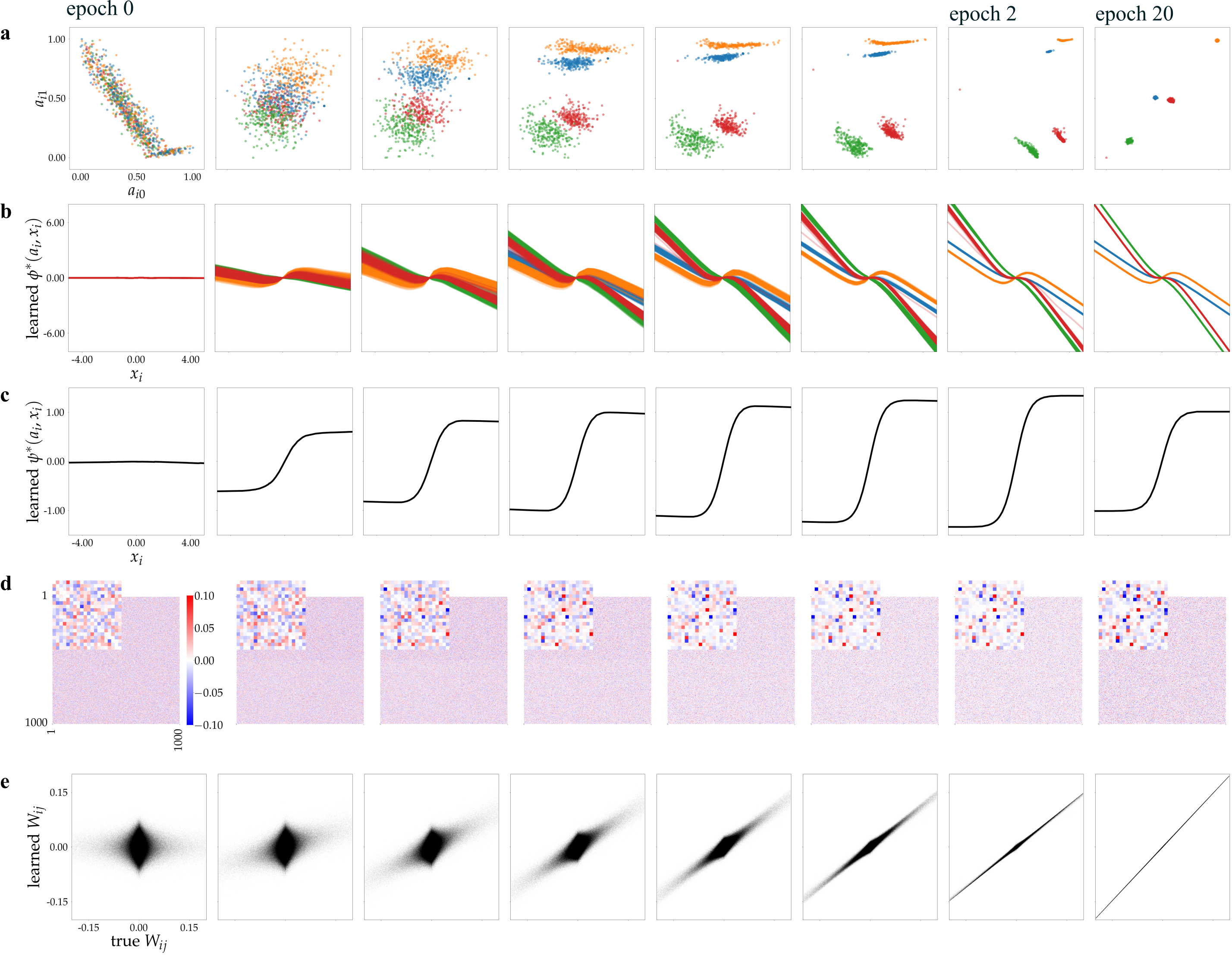}
\caption{\np{1000} densely connected neurons with 4 neuron-dependent update functions.  Results plotted over 20 epochs.  (\textbf{a}) Learned latent vectors $\Vec{a}_i$ of all neurons. (\textbf{b})~Learned update functions $\phi^*(\Vec{a}, x)$. (\textbf{c}) Learned transfer function $\psi^*(x)$, normalized to a maximum value of 1. 
(\textbf{d})~Learned connectivity $\Mat{W}_{ij}$.
(\textbf{e}) Comparison of learned and true connectivity. Colors indicate true neuron types.}
\label{supp1}
\end{figure*}

\begin{figure}[h]
\centering%
\includegraphics[width=7cm]{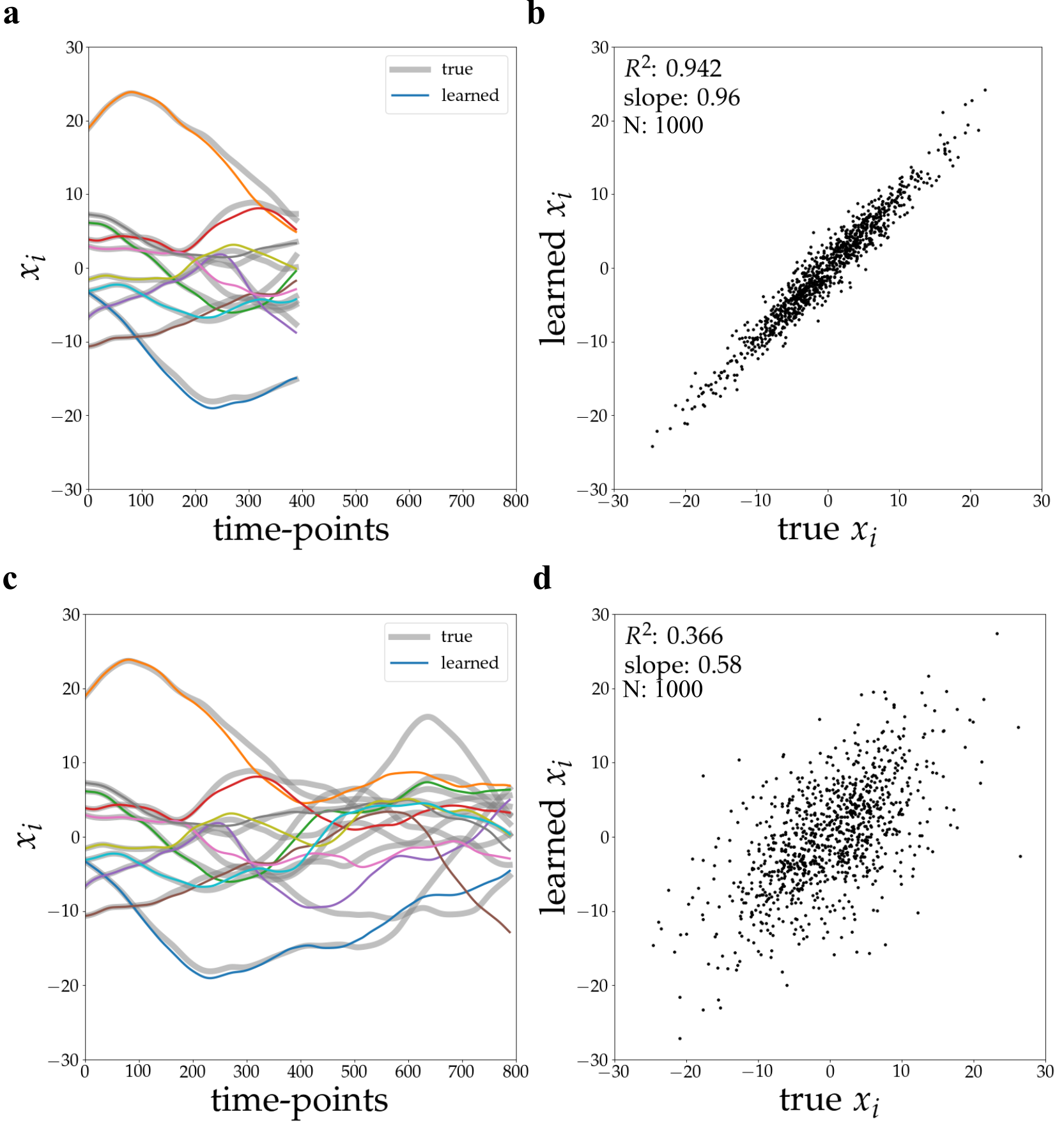}
\caption{Rollout inference performed with the GNN model trained with a simulation of \np{1000} densely connected neurons (\cref{fig2}). Results plotted at time step 400 and 800, respectively. (\textbf{a}) and (\textbf{c}) 25 learned activity traces plotted as a function of time-points. True activity traces are overlaid in light gray. (\textbf{b}) and (\textbf{d}) Comparison between true and learned activity values of 1000 neurons.  
}
\label{supp12}
\end{figure}

\begin{figure}[h]
\centering%
\includegraphics[width=\columnwidth]{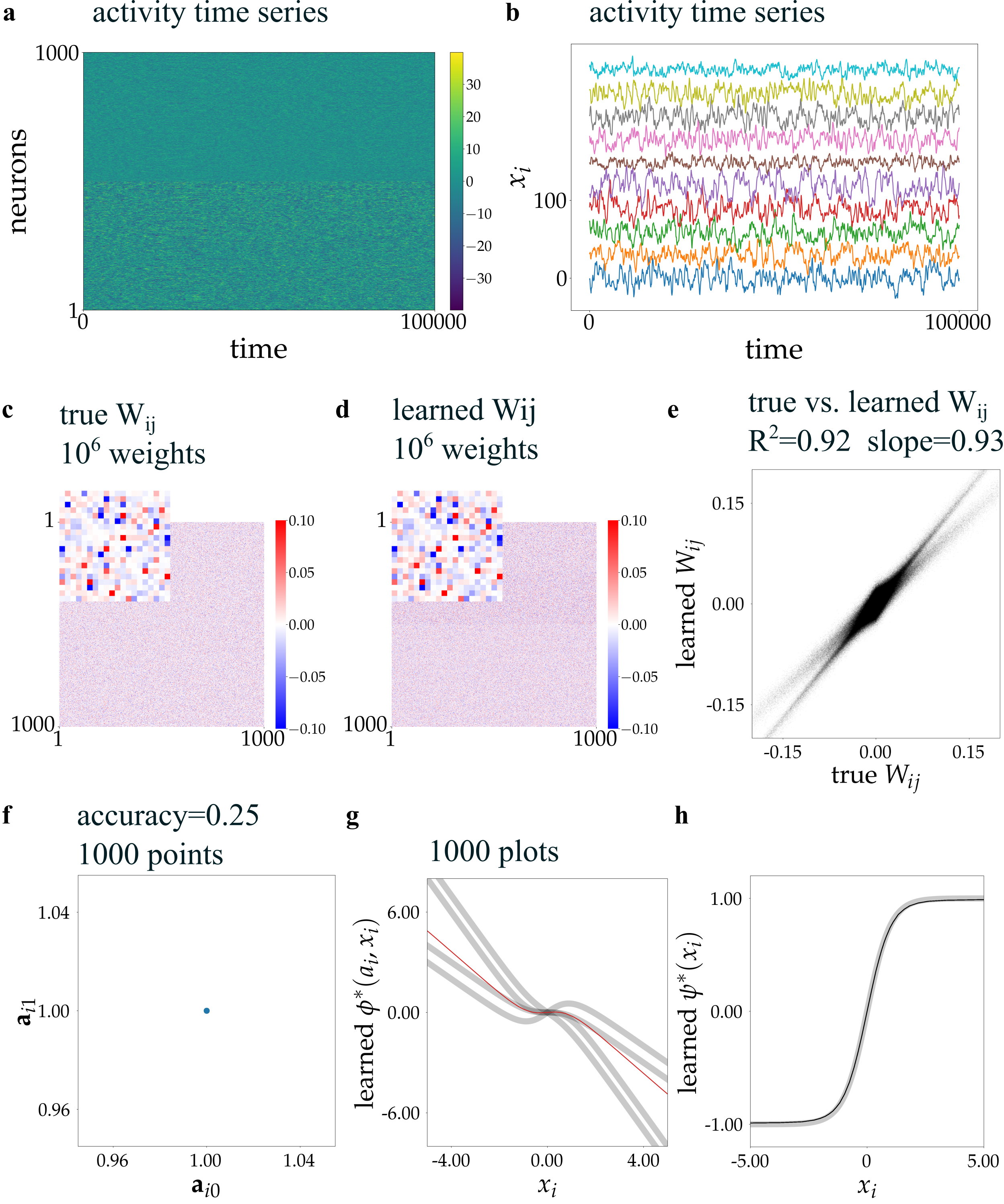}
\caption{\np{1000} densely connected neurons with 4 neuron dependent update functions (first two terms in \cref{eqn:simulation}).  Results are obtained with a GNN trained with the hypothesis that all neurons are identical. To do so, we fixed the learnable latent vectors $\Vec{a}_i$ to a unique vector. (\textbf{a}) Activity time series used for GNN training. This dataset contains $10^5$ time-points. (\textbf{b}) Sample of 10 time series taken from (\textbf{a}). (\textbf{c}) True connectivity $\Mat{W}_{ij}$. The inset shows $20\times20$ weights. (\textbf{d})~Learned connectivity. (\textbf{e}) Comparison of learned and true connectivity (given $g_i=10$ in \cref{eqn:simulation}).  (\textbf{f}) Fixed latent vectors $\Vec{a}_i$ of all neurons. (\textbf{g})~Learned update functions $\phi^*(\Vec{a}, x)$. (\textbf{h}) Learned transfer function $\psi^*(x)$,  normalized to a maximum value of 1. Colors indicate true neuron types. True functions are overlaid in light gray.}
\label{supp13}
\end{figure}

\begin{figure}[h]
\centering%
\includegraphics[width=7cm]{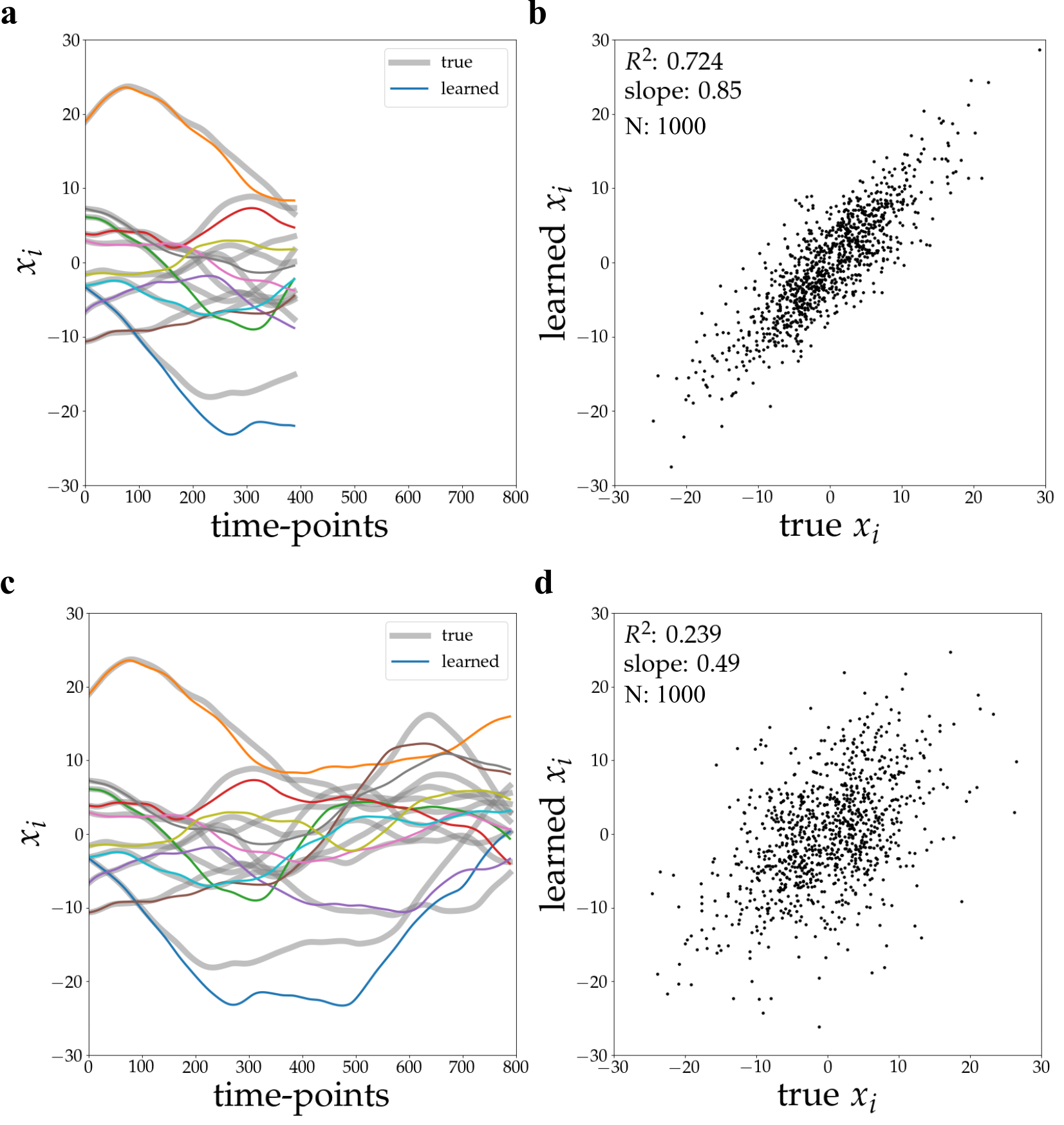}
\caption{Rollout inference performed with the GNN model trained with the hypothesis that all neurons are identical (\cref{supp13}). Results plotted at time step 200 and 800, respectively. (\textbf{a}) and (\textbf{c}) 25 learned activity traces plotted as a function of time-points. True activity traces are overlaid in light gray. (\textbf{b}) and (\textbf{d}) Comparison between true and learned activity values of 1000 neurons.  
}
\label{supp14}
\end{figure}

\begin{figure}[h]
\centering%
\includegraphics[width=7cm]{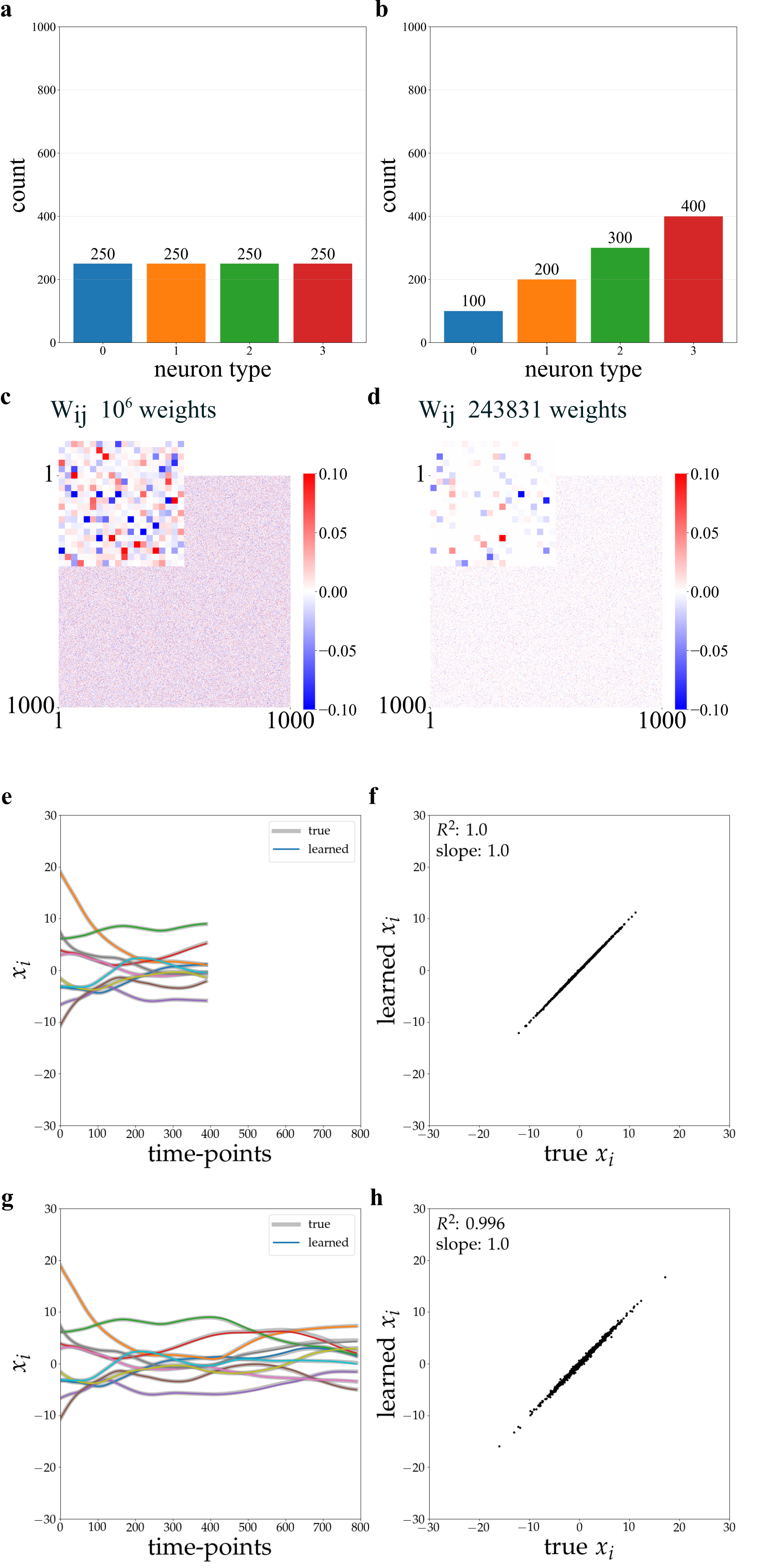}
\caption{Generalization test with modified network structure: performance evaluation after changing connectivity matrix and neuron type proportions. The GNN model was trained with \np{1000} densely connected neurons. (\textbf{a}) Original relative proportions of neuron types (25\% each). (\textbf{b}) Modified relative proportions of neuron types (10\%, 20\%, 30\%, 40\%). (\textbf{c}) Original connectivity matrix ($10^6$ weights, fully connected). (\textbf{d}) Modified sparse connectivity matrix (243,831 weights, $\sim$25\% sparsity). (\textbf{e,f}) Rollout inference over 400 time-steps shows perfect performance ($R^2=1.0$, slope$=1.0$). (\textbf{g,h}) Extended rollout over 800 time-steps maintains high accuracy ($R^2=0.996$, slope$=1.0$).}
\label{supp18}
\end{figure}

\begin{figure}[h]
\centering%
\includegraphics[width=7cm]{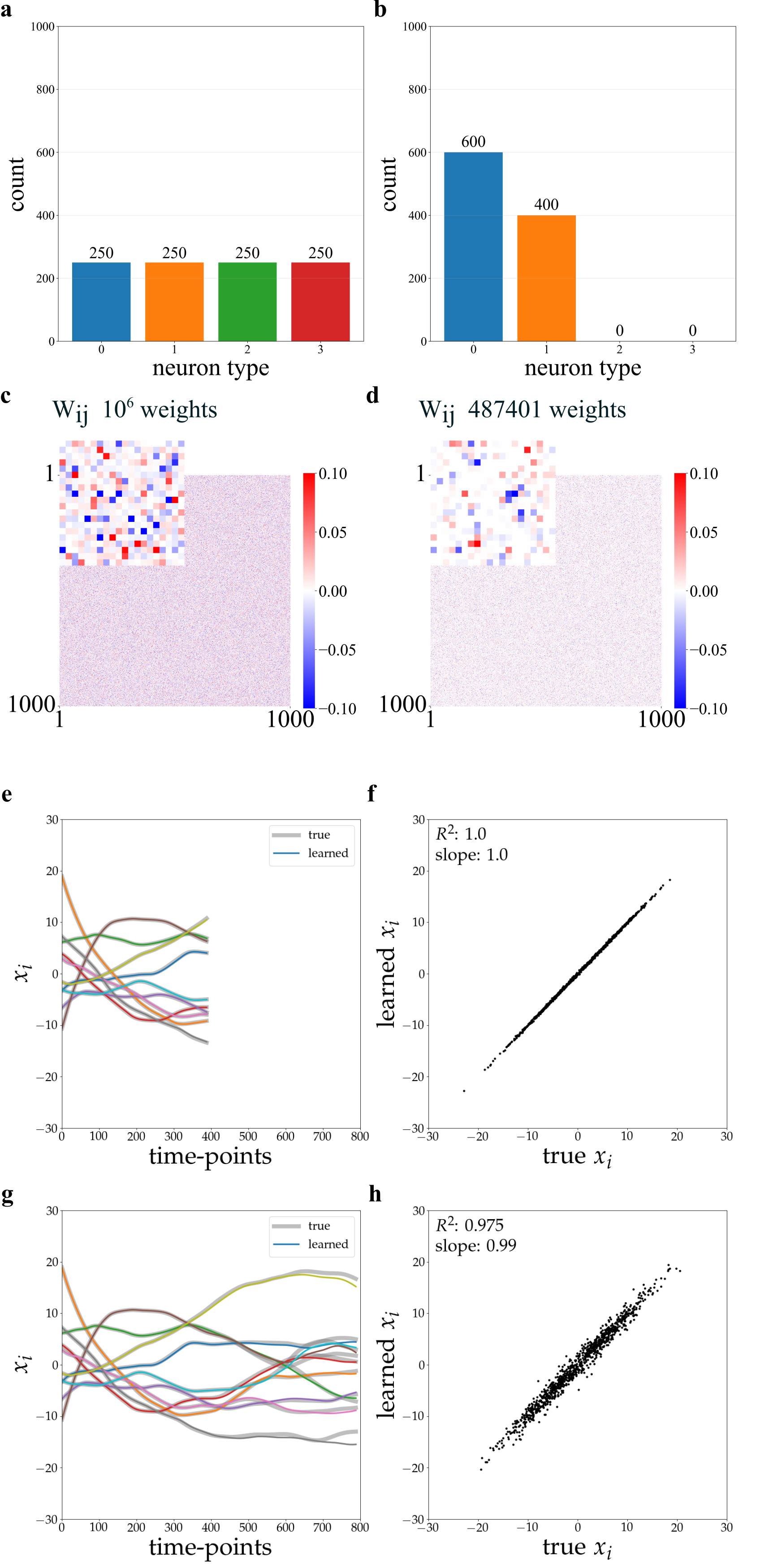}
\caption{Generalization test with modified network structure. (\textbf{a}) Original relative proportions of neuron types (25\% each). (\textbf{b}) Modified relative proportions with only two types present (60\%, 40\%, types 2 and 3 eliminated). (\textbf{c}) Original connectivity matrix ($10^6$ weights, fully connected). (\textbf{d}) Modified sparse connectivity matrix (487,401 weights, 50\% sparsity). (\textbf{e,f}) Rollout inference over 400 time-steps achieves excellent accuracy ($R^2=1.0$, slope$=1.0$). (\textbf{g,h}) Extended rollout over 800 time-steps maintains high accuracy ($R^2=0.975$, slope$=0.99$).}
\label{supp19}
\end{figure}

\begin{figure}
\centering%
\includegraphics[width=7cm]{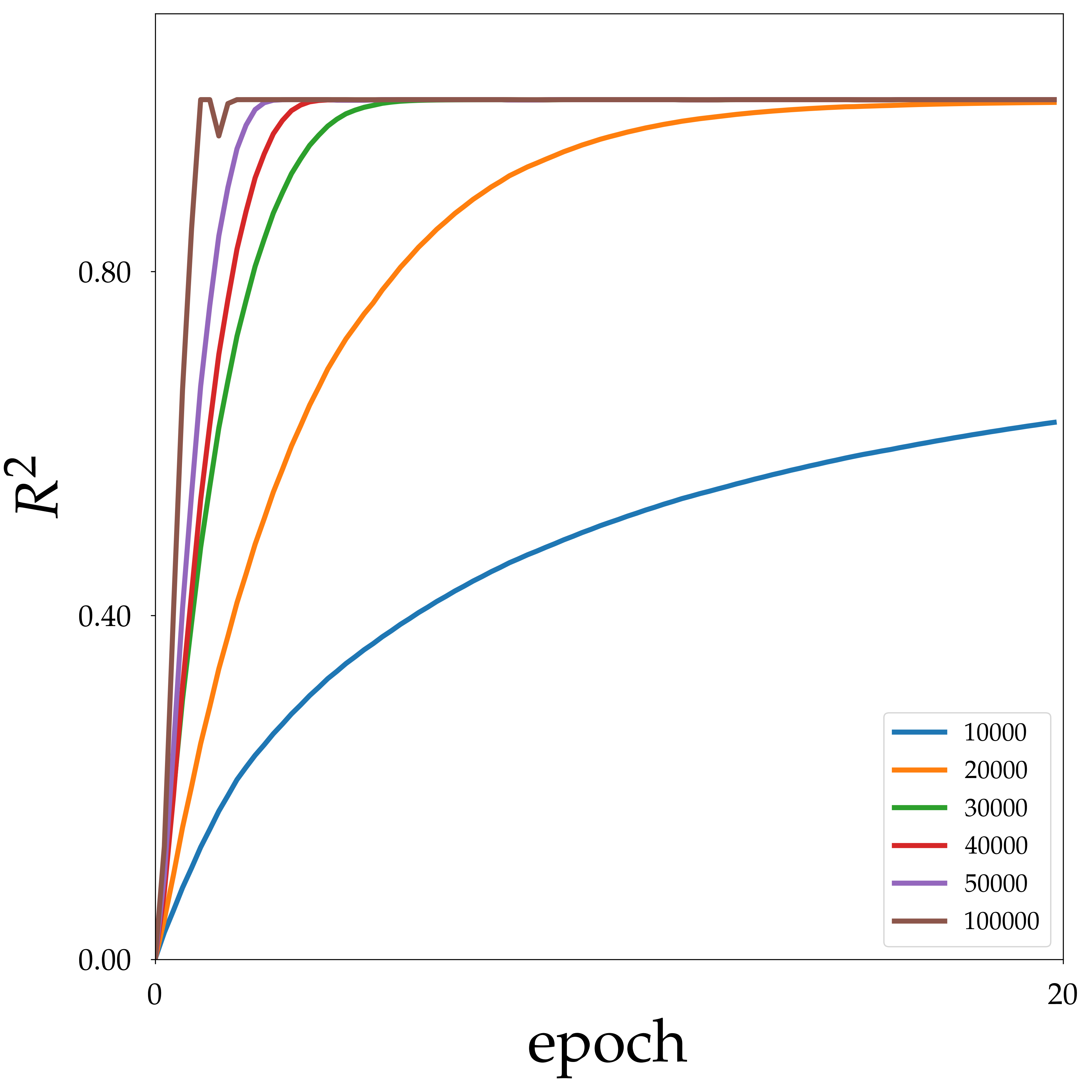}
\caption{\np{1000} densely connected neurons with 4 neuron-dependent update functions. The plot displays $R^2$ for the comparison between true and learned connectivity matrices $\Mat{W}_{ij}$ as a function of training epochs for different training dataset sizes (colors). Comparison is made at equal numbers of gradient descent iterations.}
\label{supp2}
\end{figure}

\begin{figure}
\centering%
\includegraphics[width=\columnwidth]{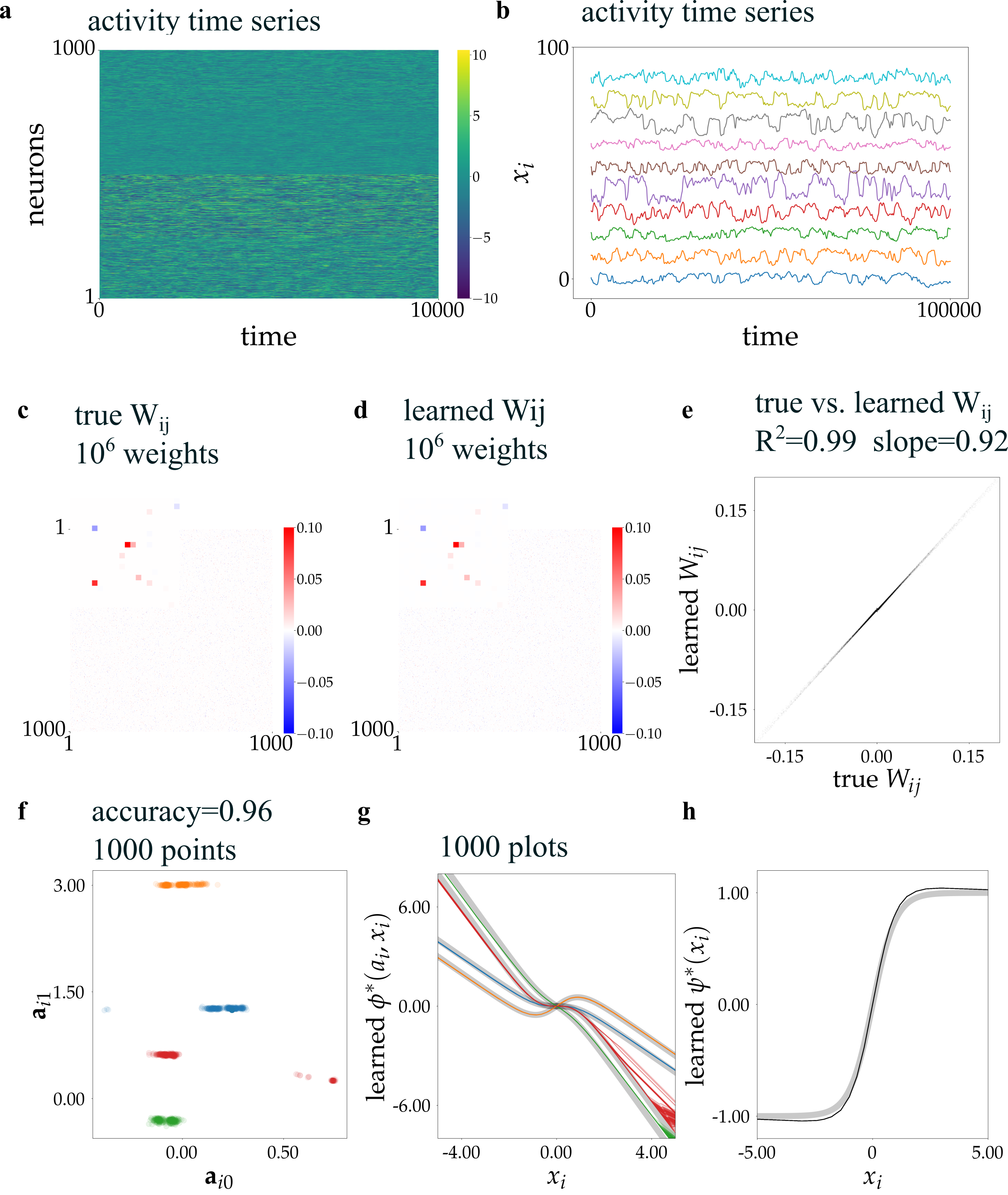}
\caption{\np{1000} sparsely (5\%) connected neurons with 4 neuron-dependent update functions. Results are obtained after 20 epochs. (\textbf{a}) Activity time series used for GNN training. The training dataset contains $10^5$ time-points. (\textbf{b}) Time series of a sample of 10 representative neurons taken from (\textbf{a}). (\textbf{c}) True connectivity $\Mat{W}_{ij}$. The inset shows $20\times20$ weights. (\textbf{d})~Learned connectivity. (\textbf{e}) Comparison of learned and true connectivity (given $g_i=10$ in \cref{eqn:simulation}). (\textbf{f}) Learned latent vectors $\Vec{a}_i$ of all neurons. (\textbf{g})~Learned update functions $\phi^*(\Vec{a}, x)$. (\textbf{h}) Learned transfer function $\psi^*(x)$,  normalized to a maximum value of 1. Colors indicate true neuron types. True functions are overlaid in light gray.}
\label{supp4}
\end{figure}

\begin{figure}
\centering%
\includegraphics[width=7cm]{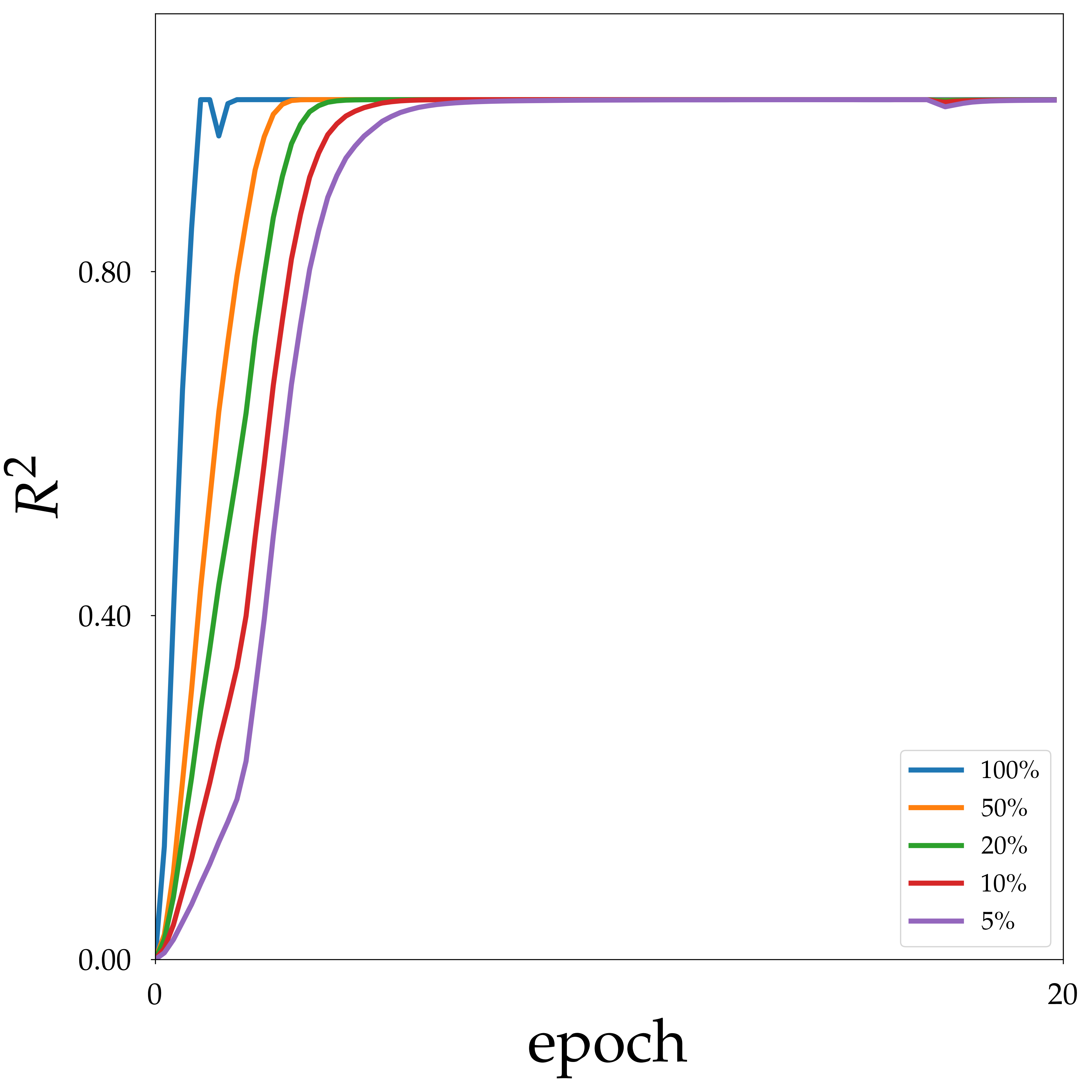}
\caption{\np{1000} densely connected neurons with 4 neuron-dependent update functions. The plot displays $R^2$ for the comparison between true and learned connectivity matrices $\Mat{W}_{ij}$ as a function of training epochs for different connectivity filling factors (colors). All comparisons are made at equal numbers of gradient descent iterations.}
\label{supp5}
\end{figure}

\begin{figure}
\centering%
\includegraphics[width=\columnwidth]{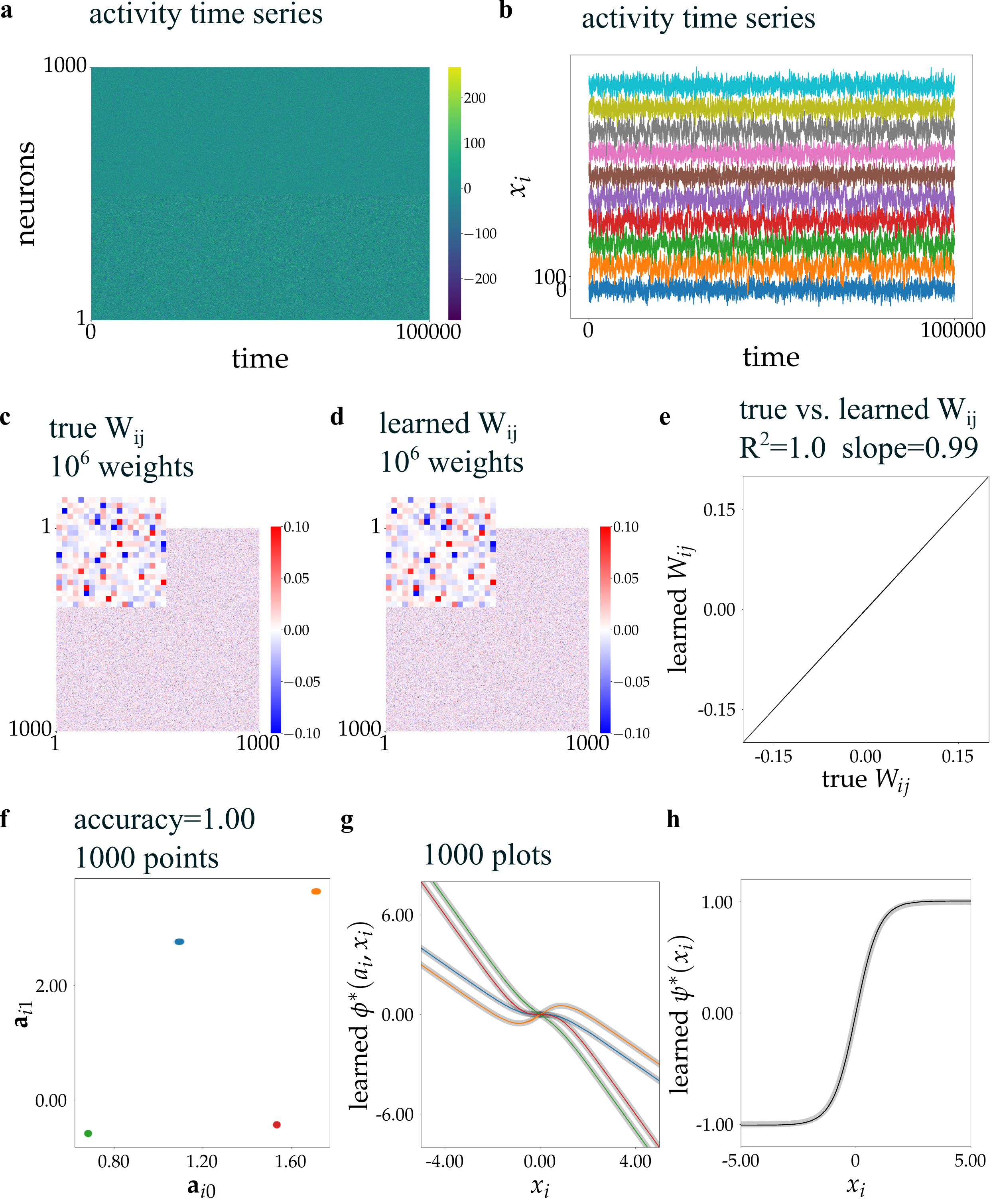}
\caption{\np{1000} densely connected neurons with 4 neuron-dependent update functions in the presence of Gaussian noise (\cref{eqn:simulation2}). The signal-to-noise ratio is about 10 dB as measured by comparing filtered and raw signals. For comparison, corresponding signals without noise are shown in \cref{fig2}. (\textbf{a}) Activity time series used for GNN training. The training dataset contains $10^5$ time-points. (\textbf{b}) Sample of 10 time series taken from (\textbf{a}). (\textbf{c}) True connectivity $\Mat{W}_{ij}$. The inset shows $20\times20$ weights. (\textbf{d})~Learned connectivity. (\textbf{e}) Comparison of learned and true connectivity (given $g_i=10$ in \cref{eqn:simulation}). (\textbf{f}) Learned latent vectors $\Vec{a}_i$ of all neurons. (\textbf{g})~Learned update functions $\phi^*(\Vec{a}, x)$. (\textbf{h}) Learned transfer function $\psi^*(x)$, normalized to a maximum value of 1. Colors indicate true neuron types. True functions are overlaid in light gray.}
\label{supp6}
\end{figure}

\begin{figure}
\centering%
\includegraphics[width=\columnwidth]{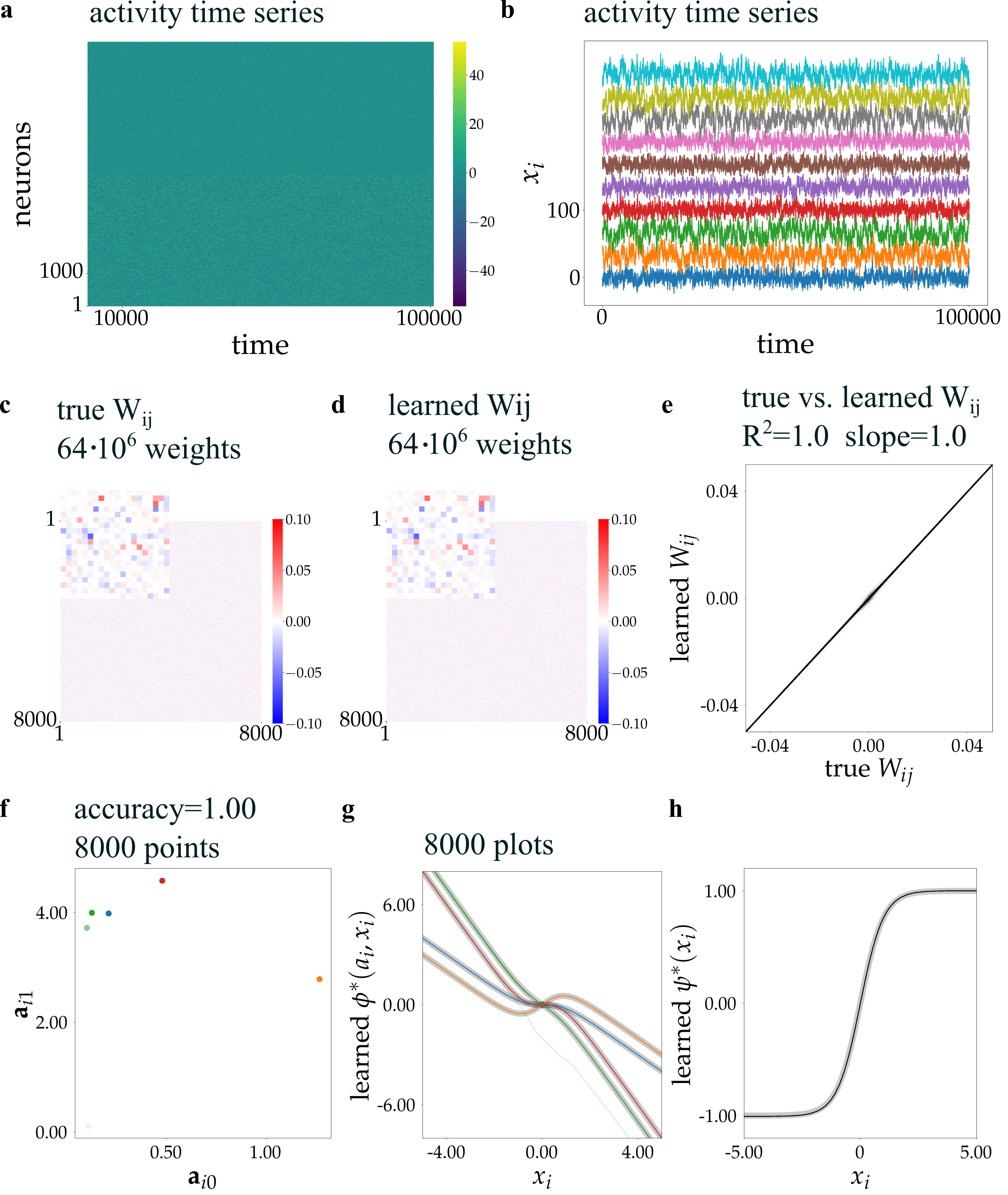}
\caption{\np{8000} densely connected neurons with 4 neuron-dependent update functions.  Results obtained after 14 epochs. (\textbf{a}) Activity time series used for GNN training. The training dataset contains $10^5$ time-points. (\textbf{b}) Sample of 10 time series taken from (\textbf{a}). (\textbf{c}) True connectivity $\Mat{W}_{ij}$. The inset shows $20\times20$ weights. (\textbf{d})~Learned connectivity. (\textbf{e}) Comparison of learned and true connectivity (given $g_i=10$ in \cref{eqn:simulation}). (\textbf{f}) Learned latent vectors $\Vec{a}_i$ of all neurons. (\textbf{g})~Learned update functions $\phi^*(\Vec{a}, x)$. (\textbf{h}) Learned transfer function $\psi^*(x)$, normalized to a maximum value of 1. Colors indicate true neuron types. True functions are overlaid in light gray.}
\label{supp3}
\end{figure}

\begin{figure}
\centering%
\includegraphics[width=\columnwidth]{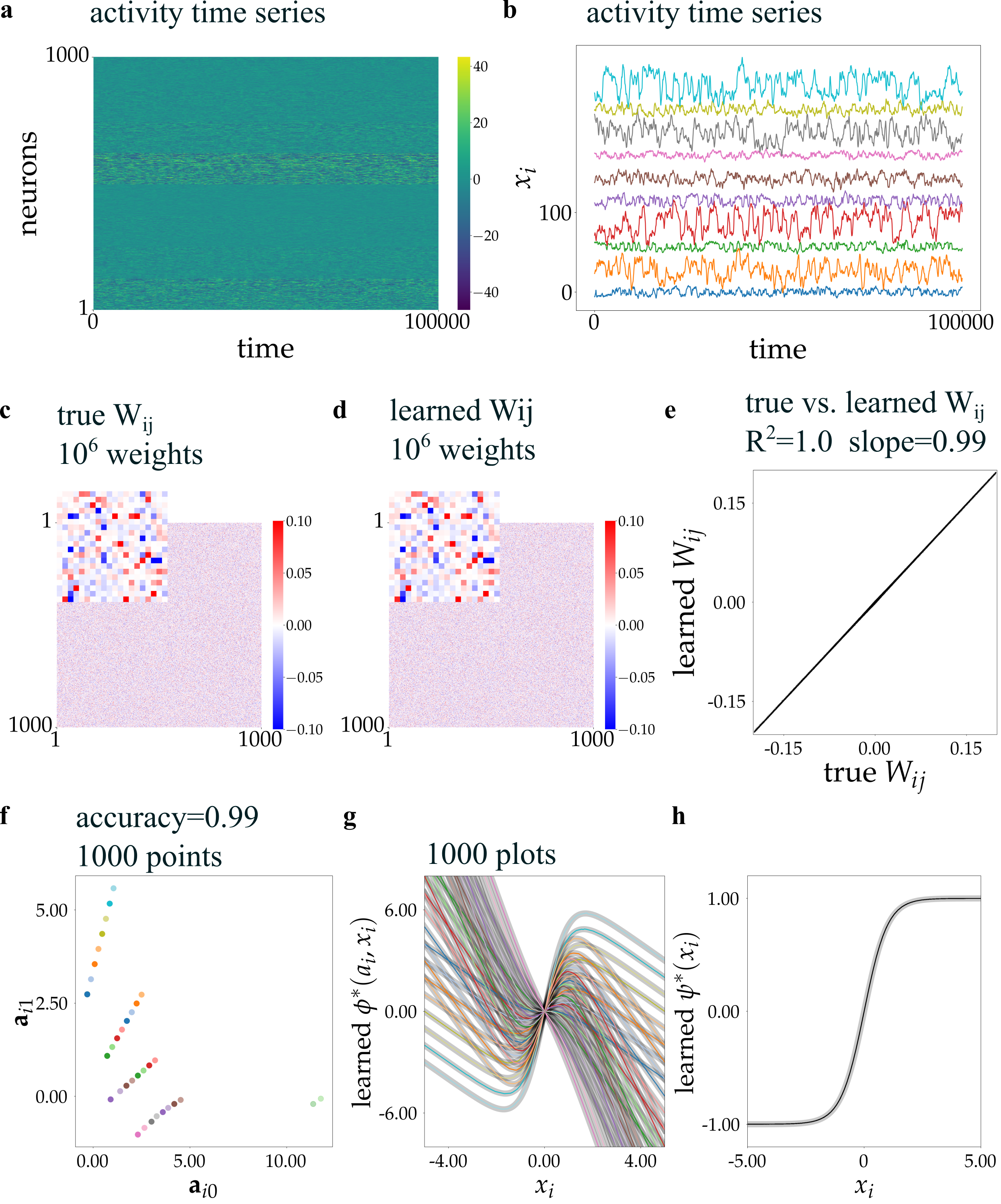}
\caption{\np{1000} densely connected neurons with 32 neuron-dependent update functions. Results are obtained after 20 epochs. (\textbf{a}) Activity time series used for GNN training. The training dataset contains $10^5$ time-points. (\textbf{b}) Sample of 10 time series taken from (\textbf{a}). (\textbf{c}) True connectivity $\Mat{W}_{ij}$. (\textbf{d}) Learned connectivity. (\textbf{e}) Comparison between learned and true connectivity. (\textbf{f}) Learned latent vectors $\Vec{a}_i$. (\textbf{g}) Learned update functions $\phi^*(\Vec{a}, x)$. (\textbf{h}) Learned transfer functions $\psi^*(x)$. Colors indicate true neuron types. True functions are overlaid in light gray.}
\label{supp7}
\end{figure}

\begin{figure}
\centering%
\includegraphics[width=\columnwidth]{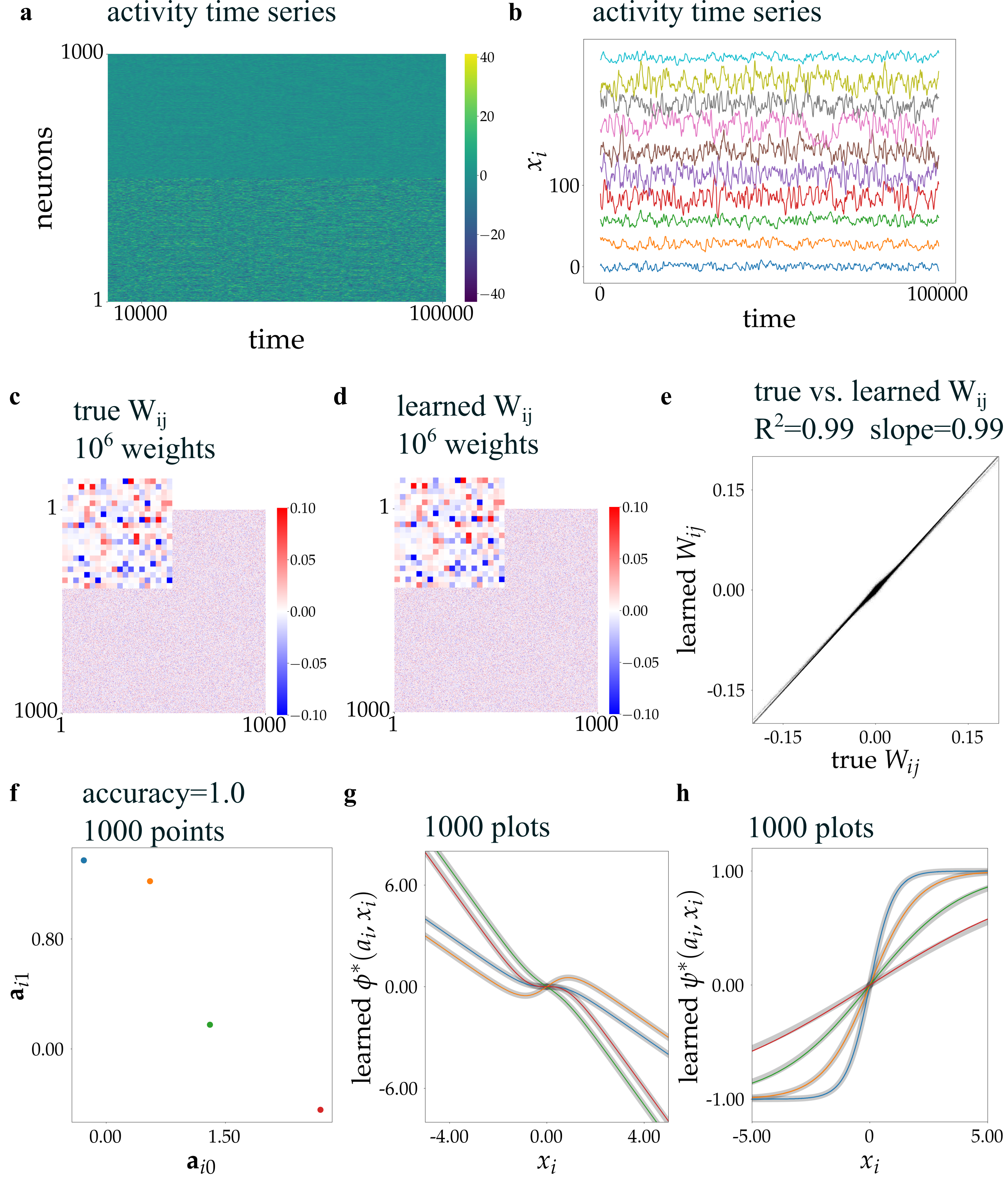}
\caption{\np{1000} densely connected neurons with neuron-dependent update and transfer functions (4 neuron types). (\textbf{a}) Activity time series used for GNN training. The training dataset contains $10^5$ time-points. (\textbf{b}) Sample of 10 time series taken from (\textbf{a}). (\textbf{c})~True connectivity $\Mat{W}_{ij}$. (\textbf{d})~Learned connectivity. (\textbf{e})~Comparison between learned and true connectivity. (\textbf{f})~Learned latent vectors $\Vec{a}_i$. (\textbf{g})~Learned update functions $\phi^*(\Vec{a}, x)$. (\textbf{h}) Learned transfer functions $\psi^*(\Vec{a},x)$. Colors indicate true neuron types. True functions are overlaid in light gray.}
\label{supp8}
\end{figure}

\begin{figure}
\centering%
\includegraphics[width=\columnwidth]{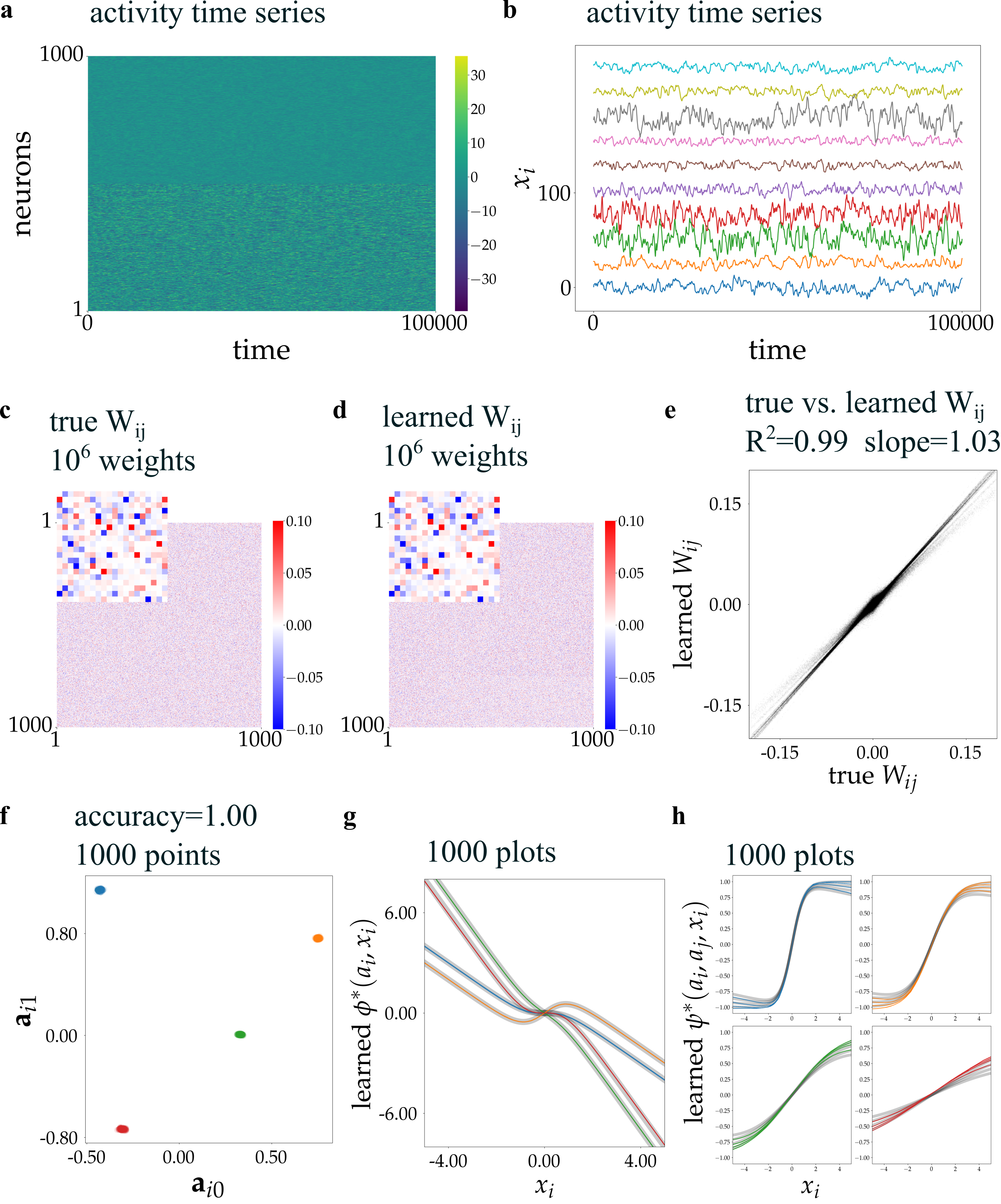}
\caption{\np{1000} densely connected neurons with neuron-dependent update and transfer functions (4 neuron types). (\textbf{a}) Activity time series used for GNN training. The training dataset contains $10^5$ time-points. (\textbf{b}) Sample of 10 time series taken from (\textbf{a}). (\textbf{c})~True connectivity $\Mat{W}_{ij}$. (\textbf{d})~Learned connectivity. (\textbf{e})~Comparison between learned and true connectivity. (\textbf{f})~Learned latent vectors $\Vec{a}_i$. (\textbf{g})~Learned update functions $\phi^*(\Vec{a}, x)$. (\textbf{h}) Learned transfer functions $\psi^*(\Vec{a_i},\Vec{a_j}, x)$. Colors indicate true neuron types. True functions are overlaid in light gray.}
\label{supp9}
\end{figure}

\begin{figure}
\centering%
\includegraphics[width=\columnwidth]{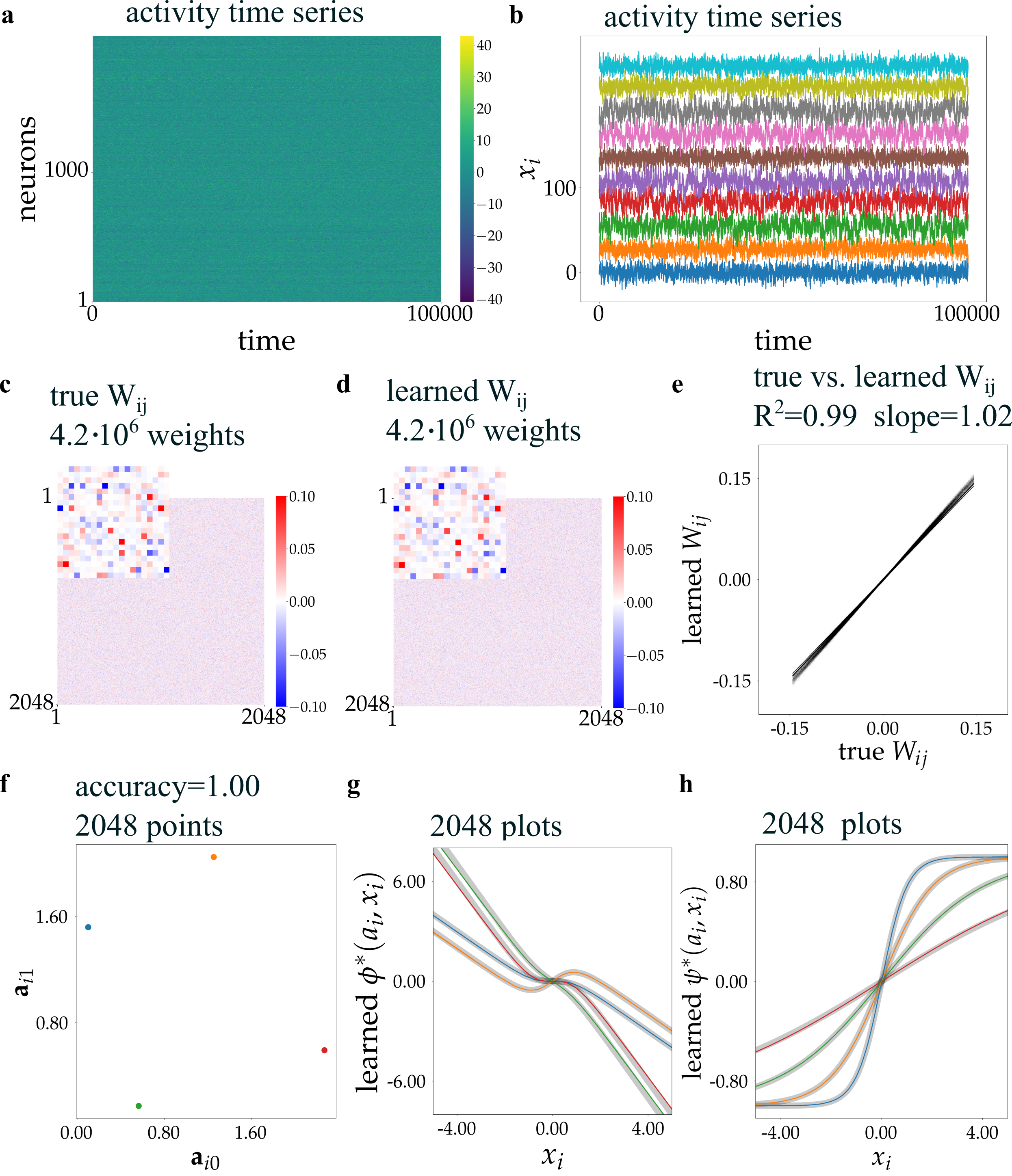}
\caption{\np{2048} densely connected neurons with neuron-dependent update and transfer functions (4 neuron types) in the presence of external stimuli. Results are obtained after 16 epochs. (\textbf{a}) Activity time series used for GNN training. The training dataset contains $10^5$ time-points. (\textbf{b}) Sample of 10 time series taken from (\textbf{a}). (\textbf{c})~True connectivity $\Mat{W}_{ij}$. (\textbf{d})~Learned connectivity. (\textbf{e})~Comparison between learned and true connectivity. (\textbf{f})~Learned latent vectors $\Vec{a}_i$. (\textbf{g})~Learned update functions $\phi^*(\Vec{a}, x)$. particleColors indicate true neuron types. True functions are overlaid in light gray.}
\label{supp10}
\end{figure}

\clearpage

\end{document}